\documentclass[10pt]{iopart}
\usepackage{iopams}
\usepackage[T2A]{fontenc}
\usepackage[utf8]{inputenc}
\usepackage{graphicx}
\usepackage{cite}
\usepackage{color}
\usepackage{multirow}
\usepackage[section]{placeins}
\usepackage{amssymb}
\expandafter\let\csname equation*\endcsname\relax
\expandafter\let\csname endequation*\endcsname\relax
\usepackage{amsmath}
\usepackage{stfloats}
\usepackage{algorithm2e}
\usepackage{url}

\makeatletter
\renewcommand{\@algocf@capt@plain}{above}
\makeatother
\SetAlFnt{\footnotesize}
\SetAlCapFnt{\small}
\SetAlCapNameFnt{\small}
\SetKwInput{KwInput}{Input}
\SetKwInput{KwOutput}{Output}

\definecolor{grey}{rgb}{0.5,0.5,0.5}
\usepackage{listings}
\lstdefinestyle{cmd}
{
	backgroundcolor=\color{grey},
	basicstyle=\scriptsize\color{white}\ttfamily
}

\begin{document}

\setlength{\parindent}{24pt}

\title[eduPIC: an introductory particle based  code for radio-frequency plasma simulation]{eduPIC: an introductory particle based code for radio-frequency plasma simulation}

\author{Zolt\'an~Donk\'o$^1$, Aranka~Derzsi$^1$, M\'at\'e~Vass$^{1,2}$, Benedek~Horv\'ath$^1$, Sebastian~Wilczek$^2$, Botond~Hartmann$^3$, Peter~Hartmann$^1$}

\address{$^1$Institute for Solid State Physics and Optics, Wigner Research Centre for Physics, 1121 Budapest, Konkoly-Thege Mikl\'os str. 29-33, Hungary\\
$^2$Department of Electrical Engineering and Information Science, Ruhr-University Bochum, D-44780, Bochum, Germany\\
$^3$\'Obudai \'Arp\'ad Secondary Grammar School, 1034 Budapest, Hungary
}

\ead{donko.zoltan@wigner.hu}

\begin{abstract}
Particle based simulations are indispensable tools for numerical studies of charged particle swarms and low-temperature plasma sources. The main advantage of such approaches is that they do not require any assumptions regarding the shape of the particle Velocity/Energy Distribution Function (VDF/EDF), but provide these basic quantities of kinetic theory as a result of the computations. Additionally, they can provide, e.g., transport coefficients, under arbitrary time and space dependence of the electric/magnetic fields. For the self-consistent description of various plasma sources operated in the low-pressure (nonlocal, kinetic) regime, the Particle-In-Cell simulation approach, combined with the Monte Carlo treatment of collision processes (PIC/MCC), has become an important tool during the past decades. 
In particular, for Radio-Frequency (RF)  Capacitively Coupled Plasma (CCP) systems PIC/MCC is perhaps the primary simulation tool these days. This approach is able to describe discharges over a wide range of operating conditions, and has largely contributed to the understanding of the physics of CCPs operating in various gases and their mixtures, in chambers with simple and complicated geometries, driven by single- and multi-frequency (tailored) waveforms. PIC/MCC simulation codes have been developed and maintained by many research groups, some of these codes are  available to the community as freeware resources. While this computational approach has already been present for a number of decades, the rapid evolution of the computing infrastructure makes it increasingly more popular and accessible, as simulations of simple systems can be executed now on personal computers or laptops. During the past few years we have experienced an increasing interest in lectures and courses dealing with the basics of particle simulations, including the PIC/MCC technique. In a response to this, this paper (i) provides a tutorial on the physical basis and the algorithms of the PIC/MCC technique and (ii) presents a basic (spatially one-dimensional) electrostatic PIC/MCC simulation code, whose source is made freely available in various programming languages. We share the code in C/C\texttt{++} versions, as well as in a version written in Rust, which is a rapidly emerging computational language. Our code intends to be a ``starting tool'' for those who are interested in learning the details of the PIC/MCC technique and would like to develop the ``skeleton'' code further, for their research purposes. Following  the description of the physical basis and the algorithms used in the code, a few examples of results obtained with this code for single- and dual-frequency CCPs in argon are also given. 
\end{abstract}

\maketitle
\ioptwocol

\section{Introduction}

\label{sec:intro}

{\it Particle based} simulations are indispensable tools for the investigation of plasma sources operated at conditions when kinetic effects prevail \cite{Basti_tutorial}. Such conditions are most prominent in weakly collisional settings, e.g., at low  pressures where the mean free path of the charged particles (typically of the electrons) can be comparable or even longer than the dimensions of the system. Under such conditions, the transport has a nonlocal character \cite{tsendin2010nonlocal,gallagher2012nonequilibrium,fu2020similarity,fu2020high,Wang2020asymmetries} and although continuum (fluid) approaches can capture this phenomenon to some extent, it is usually achieved by including the corresponding Energy/Velocity Distribution Function (EDF/VDF) and nonlocal closure of the equations as an {\it input} to the model \cite{held2004nonlocal}. Particle methods, on the other hand, do not require any assumptions of this kind and provide the EDF/VDF as a {\it result} of the calculations. The price of using particle based approaches for the description of gas discharges is their computationally intensive nature. The efficiency of the simulations can largely be enhanced by carefully designed hybrid schemes \cite{Kushner_2009,doi:10.1063/1.4922631} where different (fluid/particle) approaches are used for the different plasma species. Such approaches are needed in the presence of, e.g., a complicated plasma chemistry, which may as well necessitate simulating processes on considerably different time scales \cite{economou2017hybrid}. Improvements to fluid models can widen the range of their applicability as it was shown in \cite{becker2017advanced}, however, at low pressures particle based approaches are still expected to be more accurate.

The history of particle based simulations, including the Particle-In-Cell (PIC) approach dates back to the late 1950's when pioneering simulations \cite{Buneman_1959,Dawson_1962} have been implemented for studies of basic plasma properties and instabilities. Subsequently, for the simulation of plasma sources operating in the collisional regime, the PIC approach has been complemented with a stochastic, Monte Carlo type treatment of the collision processes, see e.g. \cite{Hockney_1988,Birdsall_2004}. This approach became known as ``Particle-In-Cell/Monte Carlo Collisions'', PIC/MCC.

We would like to note that Monte Carlo simulations of the charged particle transport in gaseous medium have been applied as well for decades, in swarm and discharge physics, e.g., \cite{lin1977monte,lin1977velocity,sakai1977development,pitchford1982comparative,boeuf1982monte,penetrante1985monte,dahl2012obtaining,grubert2013boltzmann,dujko2014monte,ponomarev2015monte}.

 While decades ago PIC and PIC/MCC codes could be executed on mainframe computers only, the advance of the computational resources has made the technique widely available by now. Today, electrostatic 1D simulations can be executed on PC-class computers and further spread of the technique is expected with the advance of Graphic Processing Units (GPUs), which considerably enhance computing performance \cite{Mertmann2011,Hanzlikova2015,shah2017novel,hur2019model,claustre2013particle,fierro2014graphics,Juhasz2020} and allow carrying out simulations efficiently in 2D and 3D as well. Consequently, the computationally intensive nature of the particle based approaches, which has been emphasized as a disadvantage for decades, is becoming less constraining these days. 

Based on the PIC/MCC approach, a wide variety of phenomena and effects have been investigated during the past decades in various plasma sources \cite{matyash2007particle}, e.g., the breakdown of the gases and the formation of the plasma \cite{radmilovic2005particle,Yongxin_breakdown}, the energy and/or angular distributions of ions at boundary surfaces \cite{georgieva2004numerical,lee2005ion,wang2007numerical,donko2018ion,o2007comparison,manuilenko2006ion}, the operation of Hall thrusters \cite{sydorenko2006kinetic,taccogna2008kinetic,lafleur2016theory}, electron heating and heating mode transitions  \cite{lafleur2014electron,gudmundsson2017electron,Liu_2015} (termed more correctly as electron power absorption and power absorption mode transitions in more recent literature \cite{Basti_tutorial,heat1,heat2}), the formation of striations \cite{PhysRevLett.116.255002}, plasma series resonance oscillations \cite{psrdonko,schungel2015self}, electron dynamics in the afterglow \cite{Proshina_2020}, as well as the effects of ion-induced \cite{daksha2017effect}, and electron-induced \cite{horvath2017eSEEmodel,horvath2018effect} secondary electrons on the plasma characteristics, the physics of fast-pulsed discharges \cite{donko2019effects} and atmospheric-pressure plasma jets \cite{korolov2019control}. Most of these investigations have found that the Electron Energy Probability Function (EEPF) in these plasmas is highly non-Maxwellian, which confirms the need for kinetic simulations.

A number of the above studies concerned low-pressure Capacitively Coupled Plasma (CCP) sources. For these, actually, the PIC/MCC method became the most widely applied technique, which proved to be indispensable in the studies of plasma sources driven by various waveforms aimed at an enhanced control of the ion properties, i.e., the Ion Flux-Energy Distribution Function (IFEDF), as well as the angular distribution of the ions at boundary surfaces. The paramount importance of these investigations stems from the applications of CCP discharges in microelectronics, photo-voltaic industry as well as in medical technologies that are based on the interaction of the active plasma species with surfaces like semiconductor wafers, medical implants, etc. \cite{Lieberman,Makabe,Pascal}. The method has frequently been applied in studies of dual-frequency RF discharges \cite{wakayama2003study,karkari2006effect}, plasma sources operated under the conditions where electrical asymmetry develops \cite{donko2008pic,zhang2011numerical}, as well as in investigations of discharges driven by Tailored Voltage Waveforms \cite{lafleur2012control,bruneau2014ion,bruneau2015effect,delattre2013radio}. Related to this, the effects of various asymmetries, like those created by a magnetic field \cite{yang2017magnetical}, by unequal secondary electron yields at the two electrodes \cite{lafleur2013secondary,korolov2013influence} have been investigated, as well as the interplay between geometrical and electrical asymmetry effects in CCPs \cite{zhang2012separate,Wang2020asymmetries}. A similarity law and the frequency scaling properties of CCPs were recently reported \cite{fu2020scaling}.

CCPs have been studied using the PIC/MCC method in various gases, e.g., in Ar \cite{braginsky2011experimental,Proshina_2020}, He \cite{sun2018pic}, O$_2$ \cite{bronold2007radio,schungel2011control,gudmundsson2013benchmark,teichmann2013particle,derzsi2015experimental,proto2018influence}, H$_2$ \cite{diomede2012hybrid}, CF$_4$ \cite{denpoh1998self,denpoh2000self,Proshina_2010,Liu_2015} and Cl \cite{huang2013particle}, as well as in Ar-CF$_4$ \cite{georgieva2003particle,brandt2019control}, Ar-O$_2$ \cite{lee2006particle}, Xe-Ar \cite{babaeva2005capacitively}, CH$_4$-H$_2$ \cite{ivanov2002comparison}, and Ar-H$_2$ \cite{voloshin_ar_H2} mixtures. The limitations and the optimization of the PIC/MCC method have been discussed, e.g, in \cite{turner2006kinetic,kawamura2000physical,Erden,sun2016pic}, and the issues of code verification and validation have been addressed in, e.g.,  \cite{turner2013simulation,turner2016verification,turner2017computer}.

The PIC/MCC method belongs to the class of particle-mesh techniques. While charged particles move in  space, their densities as well as the electric field and potential distributions (in electrostatic simulations) are computed on a discrete grid. The latter quantities are defined via the Poisson equation that takes into account the space charge generated by the ensemble of the charged particles (electrons and ions), as well as the potential distributions at the electrode surfaces that appear as boundary conditions of the differential equation. The Poisson equation has to be solved typically several thousand times within an RF cycle, at the usual ($\sim$\,10~MHz) RF frequencies. The densities of the charged particles are computed in these time steps as well, and their positions are updated, too, according to the actual value of the electric field at their positions. In the meantime, particles may reach the electrodes, where various processes (absorption, reflection, emission of additional particles)  may take place as defined by the surface model, and particles may as well collide within the gas phase with the atoms/molecules of the background gas. The above ``ingredients'' of the PIC/MCC method will be discussed in considerable detail in section \ref{sec:picmcc}.

PIC/MCC codes have been developed by many groups worldwide. Most of these serve own research purposes of the respective groups, but some are available as freely accessible useful resources, e.g. \cite{PTSG}. Excellent descriptions of the PIC/MCC method, e.g., \cite{Tskhakaya_2007,Verboncoeur_2005,kim2005particle} have been published both for electrostatic and electromagnetic cases, and for 1D, as well as for higher dimensions. Codes have been developed in a variety of languages, e.g., Pascal~\cite{Donk__2007}, C/C\texttt{++}~\cite{Verboncoeur1995,kuehn2020picfoam}, Fortran~\cite{Kolev2006,Taccogna_2013}, MATLAB~\cite{Inusa2014}, Julia~\cite{PIC_Julia}, JAVA~\cite{Markidis2005ImplementationAP}. The choice of the language can be motivated by considerations about compiler and code performance, availability of code libraries, quality of the  development and debugging environments, but primarily remains a matter of personal preference.

In this paper, we provide an elementary introduction to the PIC/MCC technique, especially for those who would like to understand every detail and are motivated to develop their ``research grade'' code from the  open source ``starting tool'' provided here in C/C\texttt{++}, as well as in Rust languages. This code is dedicated to assisting in the education of those interested in particle based plasma simulations, hence the name ``eduPIC''.
Our intention has been to keep this code as simple and as transparent as possible; optimization and further development is left for the interested readers. The codes that we provide are $\sim$ 1000 lines in length, in a single file. We include minimum diagnostics, which is, however, already sufficient for the analysis of some phenomena, as will be demonstrated in sections \ref{sec:r1} and \ref{sec:r3}. While, as it has been mentioned earlier, PIC/MCC codes are capable of addressing the physics of various plasma sources and have been aiding the understanding of various physical effects, in this paper we will apply this technique exclusively for the description of CCPs. One should, however, keep it in mind that with proper modifications, use of the  code presented here can be extended beyond CCPs.

\begin{figure}[ht ]
\begin{center}
\includegraphics[width =0.85\columnwidth]{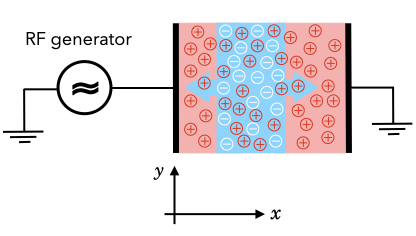}
\caption{The model system considered in the eduPIC simulation code. In experiments, the RF power is coupled to the plasma via a matching network containing a blocking capacitor. A symmetric RF discharge driven by a single-harmonic waveform, such as considered here, can be, however, modeled based on this simplified setup.}
\label{fig:model}
\end{center}
\end{figure}

Figure \ref{fig:model} presents the model system: we consider a CCP in which the electrodes are plane and parallel, and assume that the electrodes have a diameter that exceeds their separation considerably, reducing the spatial dependence of the plasma characteristics to one dimension. A rather general characteristic of such systems is that electrons are able to respond to the rapid variation of the electric field (on the nanosecond time scale), while the ions (of most gases) can only respond to the time average of the electric field \cite{chabert2020foundations}. As a consequence, the system ``splits'' into a bulk quasineutral region and sheath regions that appear/disappear near both electrodes in opposite phase during the RF cycles. The ion density distribution is therefore nearly stationary, while the electron density distribution varies significantly. Especially at low pressures, when the mean free path of the electrons may become comparable or even longer than the system size, kinetic effects are expected to arise and an accurate description of the system may be expected only from kinetic methods, like particle based simulations. Besides the electrons, the flight of the ions across the electrode sheaths may as well be strongly non-local and therefore an accurate calculation of the ion energy distribution at the surfaces also calls for a kinetic description. The PIC/MCC method clearly qualifies for these purposes, and this is clearly reflected in its increasing popularity. 

The paper is structured in the following way. Section \ref{sec:picmcc} describes the details of the PIC/MCC approach. In section \ref{sec:impl}, we discuss the details of the implementation, including the basic simulation parameters, the cross sections, thoughts on random number generation, guidance for code compilation and execution, as well as about the data files created by the code (sections \ref{sec:settings}-\ref{sec:run}). The basic (C) version of the code is explored to algorithmic details in \ref{sec:basic_code}, while a more general overview of the specific features of the C\texttt{++} and Rust codes are given in sections \ref{sec:cpp_code} and \ref{sec:rust_code}, respectively. In section \ref{sec:r1}, we present results for various discharge characteristics obtained for a set of ``reference conditions'', and in section \ref{sec:r3}, we analyze some of the characteristics of CCPs excited by a voltage waveform consisting of two components. The paper ends with a brief summary and a list of suggestions to the readers, who are interested in the further development of the eduPIC code (section \ref{sec:summ}).

\section{The principles of the PIC/MCC method}

\label{sec:picmcc}

As in CCPs the plasma density is typically in the range of $\sim 10^{9}-10^{10}$ cm$^{-3}$ and volumes of several hundreds to several thousands of cm$^{3}$ are usual, the number of electrons/ions in these systems can be in the $\sim 10^{13}$ range. Accounting for the mutual interaction and following the motion of so many particles would be impossible. Therefore, in the PIC approach, the direct (pair) interaction of the particles is omitted, the particles move under the influence of an electric field created by the potentials applied to the electrodes and modified by the space charge due to the presence of the charged particles. Additionally, instead of single particles, ``{\it superparticles}'' are traced that represent a given (high) number of individual particles. These two ideas make computations feasible with acceptable accuracy and within acceptable time.

Time is discretized into small $\Delta t = T_{\rm RF} / N_{\rm t}$ time steps, where $T_{\rm RF}$ is the period of the driving RF voltage and $N_{\rm t}$ is the number of time steps within the RF cycle. Besides time, the inter-electrode space $[0,L]$ is also discretized in the form of an equidistant computational grid having a division $\Delta x$ and $N_{\rm g}$ points. Following the C programming language convention, the time steps are numbered from 0 to $N_{\rm t}-1$, while the grid points are numbered from 0 to $N_{\rm g}-1$. The distance between the grid points is $\Delta x = L / (N_{\rm g}-1)$.  (We note, that the equidistant grid represents the simplest approach, and for specific applications, e.g., where strong field gradients are present in certain sub-domains of the computational region, a non-uniform grid can have advantages.)

The PIC/MCC method is equally suited to describe ``voltage-driven'', as well as  ``current-driven'' discharges; these cases can be realized via implementing the proper boundary conditions \cite{verboncoeur1993simultaneous}.  At this point, a note about the possibility of comparison of the simulations with experimental systems is in order. In research laboratories the electrical (voltage and current) waveforms are often measured in experiments, these make a direct comparison with simulation results possible. In industrial plasma processing, however, the main control parameter is usually the discharge power. This quantity can be computed from the simulations, but cannot be directly used as an input parameter in the form of a boundary condition. This means that one should either run simulations for several voltage values to find the conditions for a specified power, or should introduce an iteration loop into the simulation that adjusts the discharge voltage until the specified power level is reached. In our basic PIC/MCC code, we do not aim to reach this goal, the implementation of such an algorithm is left for interested readers (see also section \ref{sec:summ}).

Here, we consider the former case and, correspondingly, specify the driving voltage waveform. This waveform is directly applied to the discharge, i.e. we do not consider any external circuit. Furthermore, we assume that the electrodes are conductive. The driving voltage is applied at $x=0$ (grid point $0$) and the electrode at $x=L$ (grid point $N_{\rm g}-1$) is grounded. 

As mentioned above already, the simulation traces superparticles that represent a high number of real particles. The number of real particles represented by a superparticle is called the {\it weight} $W$, in our definition. This key factor links the number of the  superparticles with the density of the real charged particles (electrons and ions). To establish this relation, in 1D, one needs to define an arbitrary unit for the electrode surface area, which we set to $A=1$~cm$^2$. With $N_{\rm e}$ superelectrons in the simulation, we have $W\,N_{\rm e}$ real electrons in the volume $V = A\,L$. The spatially averaged electron density is $\overline{n}_{\rm e} = W\,N_{\rm e} / V$. (As an example, assuming that we have an average electron density $\overline{n}_{\rm e}=4 \times 10^{9}$ cm$^{-3}$ within an electrode gap of $L$ = 2.5 cm, at a weight of $W= 10^5$ this density is established by $N_{\rm e} = 10^5$ superelectrons.) Here, the same weight factor is used for electrons and ions. The spatially averaged ion density is $\overline{n}_{\rm i} = W\,N_{\rm i} / V$, where $N_{\rm i}$ is the number of superions in the simulation.

Finally, we note that in this simulation approach the superparticles are assumed to have a finite size, and their charge is spatially distributed according to specific cloud shapes, see, e.g., \cite{Birdsall_2004}. This finite size suppresses the short-range interaction between the particles, which can, however be considered by adding Coulomb collisions to the simulation \cite{Nanbu2000}.

In this paper, we choose a simple electropositive atomic gas (argon) as a medium for the creation of plasma and consider only a limited set of elementary processes. The charged particles that we consider are electrons and singly ionized atomic ions. This should be kept in mind during all the forthcoming discussions. One needs to note, however, that more complicated scenarios (e.g., electronegative discharges, multiple ionic species, various plasma-chemical reactions) can also be analyzed within the PIC/MCC framework, with proper extensions (not covered here).

\subsection{The basic PIC/MCC cycle}
\label{sec:six_steps}

\begin{figure}[ht ]
\begin{center}
\includegraphics[width =0.92\columnwidth]{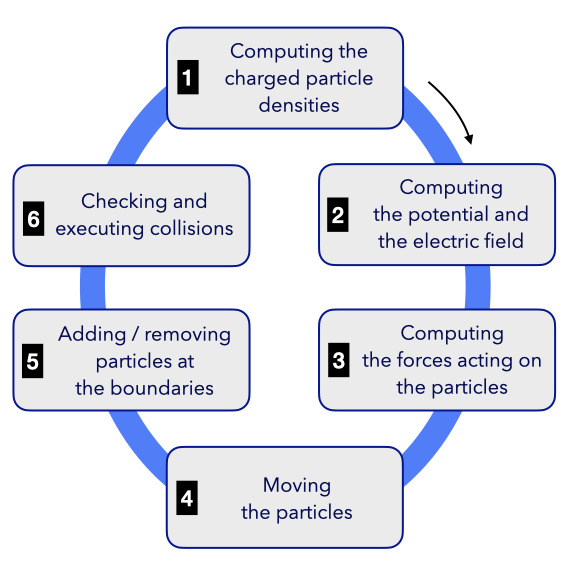}
\caption{Steps of the basic PIC/MCC cycle.}
\label{fig:pic-cycle}
\end{center}
\end{figure}

Figure~\ref{fig:pic-cycle} shows the distinct steps of the PIC/MCC cycle (which is not to be confused with the RF cycle, as the latter normally comprises thousands of the PIC/MCC cycles, given by $N_{\rm t}$), which are executed at every $\Delta t$ time step. Note, that as the motion of the ions is much slower as compared to the motion of the electrons, unequal time steps for these two species can be adopted in the simulations. This {\it ion subcycling} can result in a substantial decrease of the computational time. When this approach is used, some of the basic steps shown in figure~\ref{fig:pic-cycle} are {\it not} executed in every cycle for the ions. In the detailed description of these steps, which follows, we keep on using a ``generic'' $\Delta t$ time step. This may, however, represent different values for the electrons and the ions.   

\subsubsection{Computation of the charged particle densities}\hfill\\

\begin{figure}[ht ]
\begin{center}
\includegraphics[width =\columnwidth]{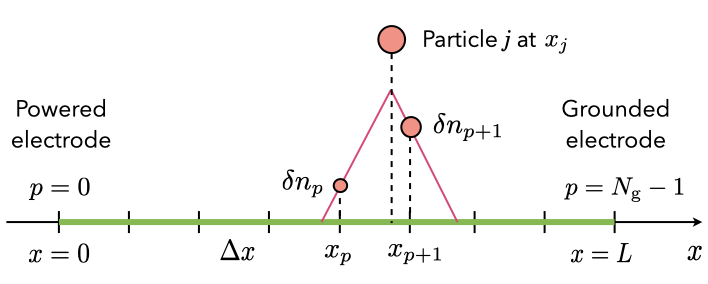}
\caption{Computation of the charged particle density at the grid points. Particle $j$ resides at position $x_j$. The green line marks the computational domain, $0 \leq p \leq N_{\rm g}-1$ is the index of the grid points.}
\label{fig:assignment}
\end{center}
\end{figure}

In this step, we determine the densities of the charged particles at the points of the computational grid. The densities at the two grid points, which enclose the particle $j$ (grid points $p$ and $p+1$, see figure \ref{fig:assignment}) are incremented by: 
\begin{equation} \label{eq:charge_assignment_1}
\delta n_p = \bigl[(p+1)-x_j/{\Delta x}\bigr] \frac{W}{A {\Delta x}},\\
\end{equation}
\begin{equation} \label{eq:charge_assignment_2}
\delta n_{p+1} = (x_j/{\Delta x} - p) \frac{W}{A {\Delta x}},
\end{equation} 
where $x_j$ is the position of the particle $j$ and $p$ is the integer part of $x_j/\Delta x$. This is done for all particles, resulting in the charge density distribution on the grid:
\begin{equation}
\rho_p = e (n_{{\rm i},p} - n_{{\rm e},p}),
\end{equation}
where $n_{{\rm i},p}$ and $n_{{\rm e},p}$ are the densities of singly charged positive ions and electrons, respectively, at grid point $p$. The above procedure actually corresponds to the concept of using particles of a finite size \cite{Birdsall_2004} (mentioned in section \ref{sec:picmcc}) - here we assume a triangular shape charge cloud.

\subsubsection{Computation of the potential and the electric field}\hfill\\

The potential distribution is obtained as the solution of the Poisson equation: 
\begin{equation} \label{eq_poisson}
    \nabla^2\phi = -\frac{\rho}{\varepsilon_0}.
\end{equation} 
This equation can be rewritten in the following finite difference form for a 1D problem:
\begin{equation} \label{eq_poisson_discrete}
    \frac{-\phi_{p-1} + 2\phi_p - \phi_{p+1}}{\Delta x^2} = \frac{\rho_p}{\varepsilon_0},
\end{equation} known as the Discrete Poisson Equation. This equation is solved on the grid: the potential distribution is calculated from the charge distribution, taking into account the potentials applied to the conductive electrodes as boundary conditions.
By differentiating the potential distribution, we obtain the electric field at each grid point: 
\begin{equation} \label{eq_efield}
    E_p = \frac{\phi_{p-1} - \phi_{p+1}}{2 \Delta x}.
\end{equation}
The boundary grid points need to be treated specially as:
\begin{equation} \label{eq_efield_b1}
    E_0 = \frac{\phi_{0} - \phi_{1}}{\Delta x} - \rho_0 \frac{\Delta x}{2 \varepsilon_0},
\end{equation}
\begin{equation} \label{eq_efield_b2}
    E_{N_{\rm g}-1} = \frac{\phi_{N_{\rm g}-2} - \phi_{N_{\rm g}-1}}{\Delta x} + \rho_{N_{\rm g}-1} \frac{\Delta x}{2 \varepsilon_0}.
\end{equation} 

\subsubsection{Computation of the forces acting on the particles}\hfill\\

\begin{figure}[ht ]
\begin{center}
\includegraphics[width =\columnwidth]{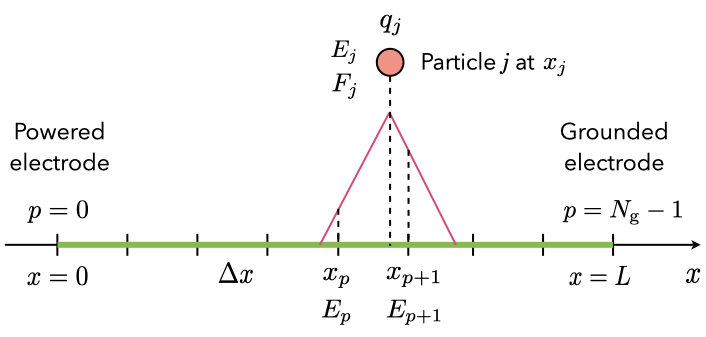}
\caption{Interpolation of the electric field (known at the grid points) to the positions of the particles. Particle $j$ with a charge $q_j$ resides at position $x_j$. The green line marks the computational domain, $0 \leq p \leq N_{\rm g}-1$ is the index of the grid points.}
\label{fig:interpolation}
\end{center}
\end{figure}

In this step, the electric field known at the grid points is interpolated to the positions of the particles. The electric field at the $x_j$ position of particle $j$, located between grid points $p$ and $p+1$ (see figure \ref{fig:interpolation}), is obtained as 
\begin{equation} \label{efield_j}
  E_j = E_p\frac{x_{p+1}-x_j}{\Delta x} + E_{p+1} \frac{x_j-x_{p}}{\Delta x},
\end{equation} 
and the force acting on the particle with a charge $q_j$ is
\begin{equation} \label{efield_j}
  F_j = E_j q_j.
\end{equation} 

\subsubsection{Moving of the particles}\hfill\\

\begin{figure}[ht ]
\begin{center}
\includegraphics[width =0.9\columnwidth]{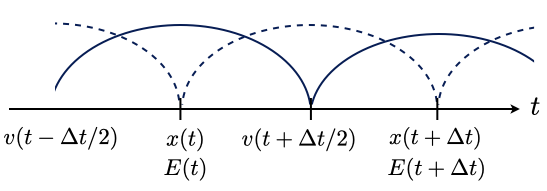}
\caption{The leapfrog integration method for advancing the particles. The particle positions ($x$) and the electric field ($E$) are known at integer time steps, while the particle velocities are defined at half integer time steps. The quantities are advanced according to eqs. (\ref{v_half_integer_t} and \ref{x_integer_t}).}
\label{fig:leapfrog}
\end{center}
\end{figure}

The particles are advanced as dictated by the equations of motion. The new positions and velocities of the particles are obtained from the solution of the discretized equation of motion. Figure~\ref{fig:leapfrog} illustrates the leapfrog integration method, where the particle velocities ($v$) are defined at half integer time steps, while the particle positions are updated at integer time steps:
 \begin{equation} \label{v_half_integer_t}
  v(t+\Delta t/2) = v(t-\Delta t/2)+\frac{q}{m}E[x(t)]\Delta t,
\end{equation} 
\begin{equation} \label{x_integer_t}
  x(t+\Delta t) = x(t)+v(t+\Delta t/2)\Delta t.
\end{equation} 

\subsubsection{Adding/removing particles at the boundaries}\hfill\\

In this step, the particles that reached the boundary surfaces are identified, and their interaction with the surfaces (e.g., absorption, reflection, secondary electron emission) is accounted for. The particles absorbed at the surfaces are removed from the simulation. The new particles created at the surfaces (e.g. secondary electrons) can be added to the simulation, according to the surface model \cite{Derzsi_2015_fastatoms, Korolov_2016_esticking, braginsky2011experimental, Hannesdottir_2016, Greb_2013,horvath2017eSEEmodel,sun2018pic,Daksha_2019,Derzsi_2020}. In the present work, we use a very simple surface model: all particles reaching the electrodes are absorbed and no particles are emitted from the electrodes.

\subsubsection{Checking and executing collisions}\hfill\\

\label{sec:collisions}

At every time step ($\Delta t$) of the given species, a decision about the occurrence of a collision has to be made for each particle. In the simplest manner, this is accomplished for each particle by (i) computing the collision probability, $P_{\rm coll}$, and (ii) comparing this probability with a random number with uniform distribution on the [0,1) interval, denoted by $R_{01}$: if $P_{\rm coll} < R_{01}$ a collision will take place. (The $R_{01}$ notation will often be used in this section, multiple occurrences of $R_{01}$ always represent different random samples of the same uniform distribution.)

The collision probability is computed as:
\begin{equation}
P_{\rm coll} = 1- \exp[-n \sigma_{\rm tot}(g) g \Delta t],
\label{eq:pcoll}
\end{equation}
where $n$ is the density of the target species, which is the background argon gas here, $\sigma_{\rm tot}$ is the {\it total cross section}, which is the sum of the cross sections of all possible collision processes of the given species, and 
\begin{equation}
    {\bf g} = {\bf v}_1 - {\bf v}_2
    \label{eq:relative_velocity}
\end{equation} 
is the {\it relative velocity} of the collision partners (${\bf v}_1$ and ${\bf v}_2$ denote the velocity of the projectile and the target particles, respectively). 

\begin{figure}[ht ]
\begin{center}
\includegraphics[width =0.9\columnwidth]{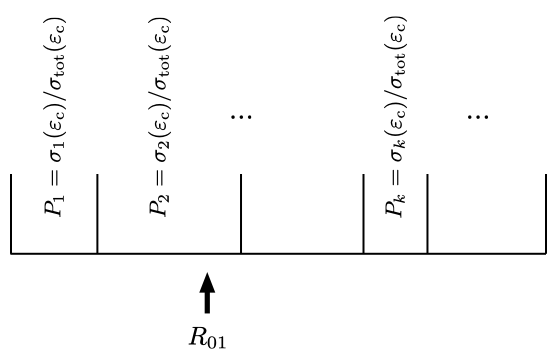}
\caption{Choosing the actual elementary process by dividing the [0,1) interval according to the probabilities ($P_k$) of occurrence of possible individual processes and selecting the type of the process by a uniformly distributed random number, $R_{01}$.}
\label{fig:choose_process}
\end{center}
\end{figure}

Whenever a collision occurs, its type is chosen in a Monte Carlo manner, by considering the values of the cross sections of all possible processes at the energy upon collision ($\varepsilon_{\rm c}= \mu g^2/2$, where $\mu$ is the reduced mass). To select the type of the collision, the ratios of the individual cross sections to the total cross section, $P_k = \sigma_k(\varepsilon_{\rm c}) / \sigma_{\rm tot}(\varepsilon_{\rm c})$, are evaluated. These values add up to 1.0, which allows dividing the [0,1) interval into segments of width $P_k$ and choosing the type of process by identifying the interval into which a random sample $R_{01}$ falls, as shown in figure \ref{fig:choose_process}. 

Using the approach outlined above, one needs to check the collision probability of {\it each} particle in {\it every} time step using the computationally expensive mathematical expression (\ref{eq:pcoll}). A more efficient selection of the colliding particles can be accomplished by the {\it null-collision} method \cite{Skullerud_1968,Nanbu2000}, where the {\it number} of colliding particles is given as $N_{\rm coll} = N P^\ast_{\rm coll}$, where $N$ is the number of particles and $P^\ast_{\rm coll} = \exp(-\nu^\ast \Delta t)$. Here, $\nu^\ast = {\rm max} \{n \sigma_{\rm tot} g\}$ is the maximum collision frequency over the domains of interest of the parameters. For the particles chosen to collide, the process is selected as above, with the addition of a null-collision process. In the case of the occurrence of this latter process, the particle proceeds without any change of its velocity. For simplicity, we do not implement this method in the eduPIC code, but suggest this to readers who would like to optimize and extend the code (see also section \ref{sec:summ}).

When the actual process is selected, the next task is to compute the post-collision velocity of the projectile. This is accomplished in several steps as follows. 

Collision events are treated in the {\it Center-of-Mass} (COM) frame. In the laboratory (LAB) frame, the center of mass moves with the velocity
\begin{equation}
    {\bf w} = \frac{m_1{\bf v}_1 + m_2{\bf v}_2}{m_1+m_2},
    \label{eq:com_velocity}
\end{equation} 
where $m_1$ and $m_2$ are the masses of the collision partners. Equations (\ref{eq:relative_velocity}) and (\ref{eq:com_velocity}) define two new variables from the ``original'' two. These equations can easily be inverted both for the pre-collision velocities and for the post-collision velocities. For us, only the latter are interesting:
\begin{equation}
{\bf v}_1' = {\bf w}' + \frac{m_2} {m_1+m_2} {\bf g}', \\
\label{eq:postcoll_velocity1} 
\end{equation}
\begin{equation}
{\bf v}_2' = {\bf w}' - \frac{m_1} {m_1+m_2} {\bf g}', \label{eq:postcoll_velocity2} 
\end{equation}
where ${\bf w}'$ and ${\bf g}'$ are the velocity of the center-of-mass and the relative velocity after the collision, respectively. The elementary classical treatment of the two-body interaction has two important conclusions: (i) due to the conservation of momentum, the velocity of the center-of-mass does not change during the encounter (i.e. ${\bf w}'={\bf w}$), and (ii) the collision results in a change of the direction of ${\bf g}$ only in the case of elastic collisions (i.e. in this case $g'=g$), while in inelastic collisions the magnitude of ${\bf g}$ changes as well due to the loss of kinetic energy.

\begin{figure}[ht ]
\begin{center}
\includegraphics[width =\columnwidth]{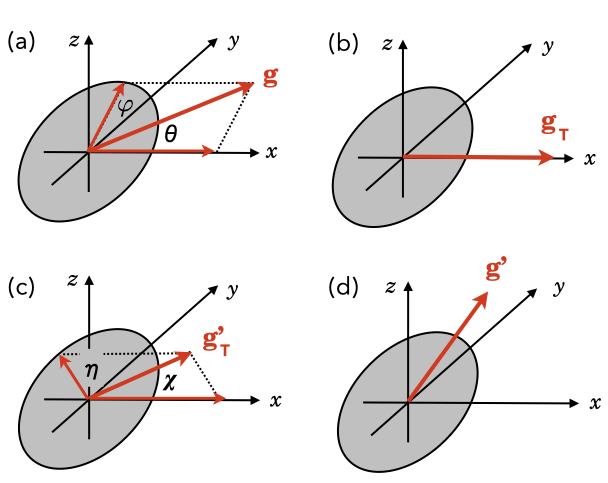}
\caption{Steps of executing the scattering process.}
\label{fig:scattering_steps}
\end{center}
\end{figure}

To execute the scattering, first we have to find the Euler angles $\theta$ and $\varphi$ that define the Cartesian components of the relative velocity ${\bf g}$ before the collision (see figure~\ref{fig:scattering_steps}(a)):
\begin{equation}
{\bf g} = {\bf v}_1 -{\bf v}_2 = \left[ \begin{array}{c}
g_x \\
g_y \\
g_z 
\end{array} \right] = g
\left[\begin{array}{c}
\cos \theta \\
\sin \theta \cos \varphi \\
\sin \theta \sin \varphi 
\end{array} \right].
\end{equation}
As the next step, we transform ${\bf g}$ to point in the $x$ direction (see figure~\ref{fig:scattering_steps}(b)); this can be achieved via two consecutive rotations of the velocity vector: (i) first, by an angle $-\varphi$ around the $x$ axis and (ii) next, by an angle $-\theta$ around the $z$ axis. Let us denote the matrices describing these operations by ${\mathbf T}_x(-\varphi)$ and ${\mathbf T}_z(-\theta)$. The transformed velocity vector is thus:
\begin{equation}
{\bf g}_{\rm T} = {\mathbf T}_z(-\theta) {\mathbf T}_x(-\varphi) {\bf g} = g
\left[ \begin{array}{c}
1 \\
0 \\
0 
\end{array} \right].
\end{equation}
Actually, this transformation does not need to be carried out, as one can {\it assume} that ${\bf g}_{\rm T}$ points in the $x$ direction. The inverse of this transformation will, however, be needed later on. 

The deflection and the change of the magnitude of the velocity vector is defined by the type of the collision. In the present model, we consider elastic, excitation and ionization collisions for the electrons, as well as elastic collisions for the ions. The scattering of the electrons is assumed to be isotropic in all types of processes, while for the ions a differential cross section that comprises an isotropic and a backward scattering part is adopted, as advised in~\cite{phelps1994application}.

In elastic collisions (both e$^-$+Ar and Ar$^+$+Ar), the magnitude of ${\bf g}$ is unchanged as dictated by the conservation of the total momentum. In this case, it is only the direction of ${\bf g}$ that has to be changed. 

The deflection of ${\bf g}$ is defined by two angles: the scattering angle $\chi$, and the azimuth angle $\eta$. The former has to be set according to the differential cross section of the scattering process, while the latter has a uniform distribution over the $[0,2\pi)$ interval. These angles, in accordance with the Monte Carlo approach, are generated based on random numbers. For isotropic scattering,  we compute $\chi$ as:
\begin{equation}
\chi = \arccos (1- 2 R_{01}),
\end{equation}
while for backward scattering:
\begin{equation}
\chi = \pi.
\end{equation}
The azimuth angle is set in all cases to be:
\begin{equation}
\eta = R_{01} 2 \pi.
\end{equation}

In case of inelastic collisions (that we consider for electrons only), the magnitude of the velocity vector has to be changed according to: 
\begin{equation}
\frac{1}{2} \mu (g')^2 = \frac{1}{2} \mu g^2 - \Delta \varepsilon,
\end{equation}
where $\Delta \varepsilon$ is the excitation or ionization energy for the actual process and $\mu$ is the reduced mass of the projectile/target system. 

In the case of ionization ($\Delta \varepsilon = \varepsilon_{\rm ion}$), the ``scattered'' and ``ejected'' electrons share the remaining energy:
\begin{equation}
\varepsilon_{\rm s} + \varepsilon_{\rm e} = \varepsilon - \varepsilon_{\rm ion},
\end{equation} 
where $\varepsilon$ is the kinetic energy of the original projectile electron, $\varepsilon_{\rm s}$ and $\varepsilon_{\rm e}$, respectively, are the energies of the scattered and ejected electrons. The partitioning between these latter two can be based on a distribution derived from experimental data \cite{Opal}:
\begin{equation}
\varepsilon_{\rm e} = w \tan \biggl[ R_{01} \arctan \Bigl( \frac{\varepsilon - \varepsilon_{\rm ion}}{2 w} \Bigr) \biggr],
\end{equation}
where $w$ is a characteristic parameter of the gas, in the case of Ar, $w=10$. The scattering angles of the electrons are computed from
\begin{equation}
    \chi_{\rm s} = \sqrt{\varepsilon_{\rm s}/(\varepsilon-\varepsilon_{\rm ion}})~,~
    \chi_{\rm e} = \sqrt{\varepsilon_{\rm e}/(\varepsilon-\varepsilon_{\rm ion})},
\end{equation}
whereas the azimuth angles are generated as
\begin{equation}
    \varphi_{\rm s} = 2\pi R_{01}~,~
    \varphi_{\rm e} = \varphi_{\rm s} + \pi.
\end{equation}

Having generated the scattering and azimuth angles, and having as well changed the magnitude of the velocity ($g_{\rm T} \rightarrow g_{\rm T}'$) due to the energy change, the velocity vector is deflected, as shown in figure~\ref{fig:scattering_steps}(c): 
\begin{equation}
{\bf g}_{\rm T}' = g'
\left[ \begin{array}{c}
\cos \chi \\
\sin \chi \cos \eta \\
\sin \chi \sin \eta 
\end{array} \right],
\end{equation}
and is subsequently transformed back to the original coordinate system by the inverse rotation operations (see figure~\ref{fig:scattering_steps}(d)):
\begin{eqnarray}
{\bf g}' = {\mathbf T}_x(\varphi) {\mathbf T}_z(\theta) {\bf g}_{\rm T}' = \\ \nonumber
g' \left[ \begin{array}{ccc}
\cos \theta & -\sin \theta & 0 \\
\sin \theta \cos \varphi & \cos \theta \cos \varphi & -\sin \varphi \\
\sin \theta \sin \varphi & \cos \theta \sin \varphi & \cos \varphi  
\end{array} \right] 
\left[ \begin{array}{c}
\cos \chi \\
\sin \chi \cos \eta \\
\sin \chi \sin \eta 
\end{array} \right],
\end{eqnarray}
which gives the final result for the post-collision velocity vector as:
\begin{eqnarray}
{\bf v}_1' = {\bf w} + \frac{m_2}{m_1+m_2} g' \times \\ \nonumber 
\resizebox{\columnwidth}{!}{\begin{math}
\left[ \begin{array}{c}
\cos \theta \cos \chi - \sin \theta \sin \chi \cos \eta \\ 
\sin \theta \cos \varphi \cos \chi + \cos \theta \cos \varphi \sin \chi \cos \eta  - \sin \varphi \sin \chi \sin \eta  \\
\sin \theta \sin \varphi \cos \chi + \cos \theta \sin \varphi \sin \chi \cos \eta + \cos \varphi \sin \chi \sin \eta  
\end{array} \right].\end{math}}
\end{eqnarray}

Finally, we note that for the e$^-$+Ar collision we adopt the {\it cold gas approximation}, i.e., assume that the target Ar atoms are at rest. Correspondingly, ${\bf v}_2$ in eq. (\ref{eq:relative_velocity}) is set to zero. For the Ar$^+$+Ar collisions, on the other hand, a random potential collision partner is sampled from the thermal distribution of the Ar atoms, before computing the collision probability. This sampling is aided by a random variable with normal distribution.

\subsection{Accuracy and stability criteria}

\label{sec:stability}

In order to have a stable simulation and to ensure that the results are accurate, a number of accuracy/stability conditions, which depend on the simulation settings, have to be fulfilled (see e.g. \cite{kim2005particle,lymberopoulos1995two}). These conditions are as follows:
\begin{enumerate}
    \item {\it The spatial grid has to resolve the electron Debye length,} i.e., $\Delta x / \lambda_{\rm D} \lesssim 1$ should hold, where the Debye length $\lambda_{\rm D} = \sqrt{\varepsilon_0 k_{\rm B} T_{\rm e}/ n_{\rm e} e^2}$ is evaluated with the effective electron temperature derived, e.g., from the mean electron energy;
    \item {\it The integration time step of the equations of motion has to be small enough to ensure that trajectories are computed accurately,} i.e., $\omega_{\rm p,e} \Delta t_{\rm e} <2.0$ should hold. This condition is usually more restrictive for electrons, as electron oscillations represent the fastest movement in the system. In practice, a factor of 10 smaller time step is routinely used;
    \item {\it The collision probabilities should be sufficiently small}, in order to minimize the effect of ``missed'' collisions (i.e. multiple collisions during a time step of the given species). It is a good practice to keep the collision probability of a given species (given in equation (\ref{eq:pcoll})) below $P_{\rm coll} \cong 0.05$;
    \item {\it Particles should not fly a longer distance during a time step than the grid division, as expressed by the Courant–Friedrichs–Lewy condition} \cite{karimabadi2005new}. This condition states that $v_{\rm max} \Delta t / \Delta x < 1$ should hold, where $v_{\rm max}$ is the maximum velocity of the particles. In practice, we loosen this condition by requiring it to hold up to a value $v_{\rm max}^\ast < v_{\rm max}$, where $v_{\rm max}^\ast$ corresponds to the energy, where the EEPF decays to a marginal value;
    \item {\it Accurate results require a high number of particles per grid cell.} Artificial numerical heating \cite{turner2006kinetic} appears in the system due to fluctuations, which are stronger when a smaller number of (super)particles is used. In an ideal PIC/MCC simulation, the results should not depend on the particle number (or on the weight, $W$). However, due to the above mentioned effect, this is not exactly the case, see e.g. \cite{Erden,sun2016pic}.  
\end{enumerate}

In practice, the first condition dictates the number of grid points, while the next three criteria define the magnitudes of the time steps for the electrons and ions. Which of them is most restrictive depends on the operating conditions. 

The last condition in the list represents a real bottleneck in the PIC/MCC approach, the choice of the number of particles is always a compromise between accuracy and the execution time of the computer program.

\section{Implementations}

\label{sec:impl}

In the following subsections, we provide details about the basic simulation parameters (section \ref{sec:settings}), the cross sections used (section \ref{sec:cs}), random number generation (section \ref{sec:rng}), as well as about code compilation and execution (sections \ref{sec:CComp} and \ref{sec:run}, respectively). In section \ref{sec:basic_code}, a detailed explanation of the structure of the ``basic'' version of the code will be given, which has been written in C, with minimum C\texttt{++} extensions. The latter include the use of constant declarations and the utilization of the random number generators from the C\texttt{++} standard library. Despite the use of these C\texttt{++} features, we refer to the basic code as a ``C code'' throughout this paper. As already stated earlier, this version of the code is optimized for transparency and easy readability. Subsequently, more sophisticated versions of the code are also discussed, in sections \ref{sec:cpp_code} and \ref{sec:rust_code}, which have been written, respectively, in C\texttt{++} and Rust languages. 

When discussing the implementation details of the PIC/MCC approach in the eduPIC code we use the notations for the quantities as they appear in the codes. Table 1 presents a ``dictionary'' between these notations and those used in the previous parts of the paper. 

\begin{table*} [tp]
\centering%
\caption{Dictionary of the notations of some physical quantities and simulation parameters, which appear in the theoretical part (section \ref{sec:picmcc}) and in the parts of the code discussed in this section. (Following the conventions of the C language, constants appear in capital letters.)}
\begin{tabular}{lll}
\hline
Quantity & Notation & In the code \\ 
\hline
{\it Constants:}\\
Voltage amplitude & $V_0$ & \texttt{VOLTAGE} \\
Excitation frequency & $f$ & \texttt{FREQUENCY} \\
Electrode gap & $L$ & \texttt{L} \\
Number of grid points & $N_{\rm g}$ & \texttt{N\_G} \\
Division of spatial grid & $\Delta x$ & \texttt{DX} \\
Number of time steps & $N_{\rm t}$ & \texttt{N\_T} \\
Electron time step & $\Delta t_{\rm e}$ & \texttt{DT\_E} \\
Ion time step & $\Delta t_{\rm i}$ & \texttt{DT\_I} \\
Subcycling ratio & $N_{\rm s}$ & \texttt{N\_SUB} \\
Superparticle weight & $W$ & \texttt{WEIGHT} \\
Number of entries in cross section tables &  & \texttt{CS\_RANGES }\\
Energy resolution of cross section tables &  & \texttt{DE\_CS }\\
Energy resolution of the EEPF &  & \texttt{DE\_EEPF}\\
Energy resolution of the IFEDF &  & \texttt{DE\_IFED}\\
Size of particle coordinate arrays &  & \texttt{MAX\_N\_P} \\
\hline
{\it Variables:}\\
Potential & $\Phi(x)$ & \texttt{pot[ ]} \\
Electric field & $E(x)$ & \texttt{efield[ ]}  \\
Electric charge density & $\rho(x)$ & \texttt{rho[ ]}\\
Number of electrons & $N_{\rm e}$ & \texttt{N\_e} \\
Positions of electrons & $x_{\rm e}$ & \texttt{x\_e[ ]}\\
Electron velocity vector & ${\bf v}_{\rm e}$ & \texttt{vx\_e[ ],vy\_e[ ],vz\_e[ ]} \\
Electron density & $n_{\rm e}(x)$ & \texttt{e\_density[ ]} \\
Number of positive ions & $N_{\rm i}$ & \texttt{N\_i} \\
Positions of ions & $x_{\rm i}$ & \texttt{x\_i[ ]}\\
Ion velocity vector& ${\bf v}_{\rm i}$ & \texttt{vx\_i[ ],vy\_i[ ],vz\_i[ ]} \\
Positive ion density & $n_{\rm i}(x)$ & \texttt{i\_density[ ]} \\
Electron impact total cross section & $\sigma_{\rm e}(\varepsilon)$ & \texttt{sigma\_tot\_e[ ]} \\
Ion impact total cross section & $\sigma_{\rm i}(\varepsilon)$ & \texttt{sigma\_tot\_i[ ]} \\
\hline
\end{tabular}
\label{table1}
\end{table*}

\subsection{Basic simulation settings} 
\label{sec:settings}

The code simulates a CCP with conducting electrodes, placed at a distance \texttt{L} from each other. The space between them is filled with the background gas that has a spatially uniform density defined by the gas pressure (\texttt{PRESSURE}) and the temperature (\texttt{TEMPERATURE}).
The electrode gap, including the boundaries, comprises \texttt{N\_G} grid points, the spacing between the grid points is \texttt{DX}. For computational reasons, the inverse of this quantity, \texttt{INV\_DX} is also defined, and in cases when a division by \texttt{DX} is required in the code, a more efficient multiplication operation by \texttt{INV\_DX} is executed instead.  

By default, one of the electrodes is driven by a cosine waveform characterised by the parameters \texttt{VOLTAGE} and \texttt{FREQUENCY}. The excitation waveform is specified within the \texttt{solve\_Poisson()} function. One period of the RF excitation is divided into \texttt{N\_T} time steps of length \texttt{DT\_E}, which is the base time step for handling electrons. A longer time step \texttt{DT\_I} is defined for the ions via the subcycling parameter \texttt{N\_SUB}.

All input parameters (specifying the geometry, the driving waveform, the time steps, etc.), as well as all the physical constants are specified in SI units in the code, with the exception of the constants that define the energy resolution of the EEPF and the IFEDF.

Particle coordinates reside in static arrays. For the electrons, e.g., \texttt{x\_e[ ]} stores the positions, while \texttt{vx\_e[ ]}, \texttt{vy\_e[ ]}, and \texttt{vz\_e[ ]} store the Cartesian velocity components. The size of these arrays in the code is set by the constant \texttt{MAX\_N\_P}, which has a predefined value of  1\,000\,000, which is likely to be sufficient for applications of this code. Similar arrays are used for the coordinates of the ions.

\subsection{Cross sections} 
\label{sec:cs}

\begin{figure}[ht]
\begin{center}
\includegraphics[width =0.5\textwidth]{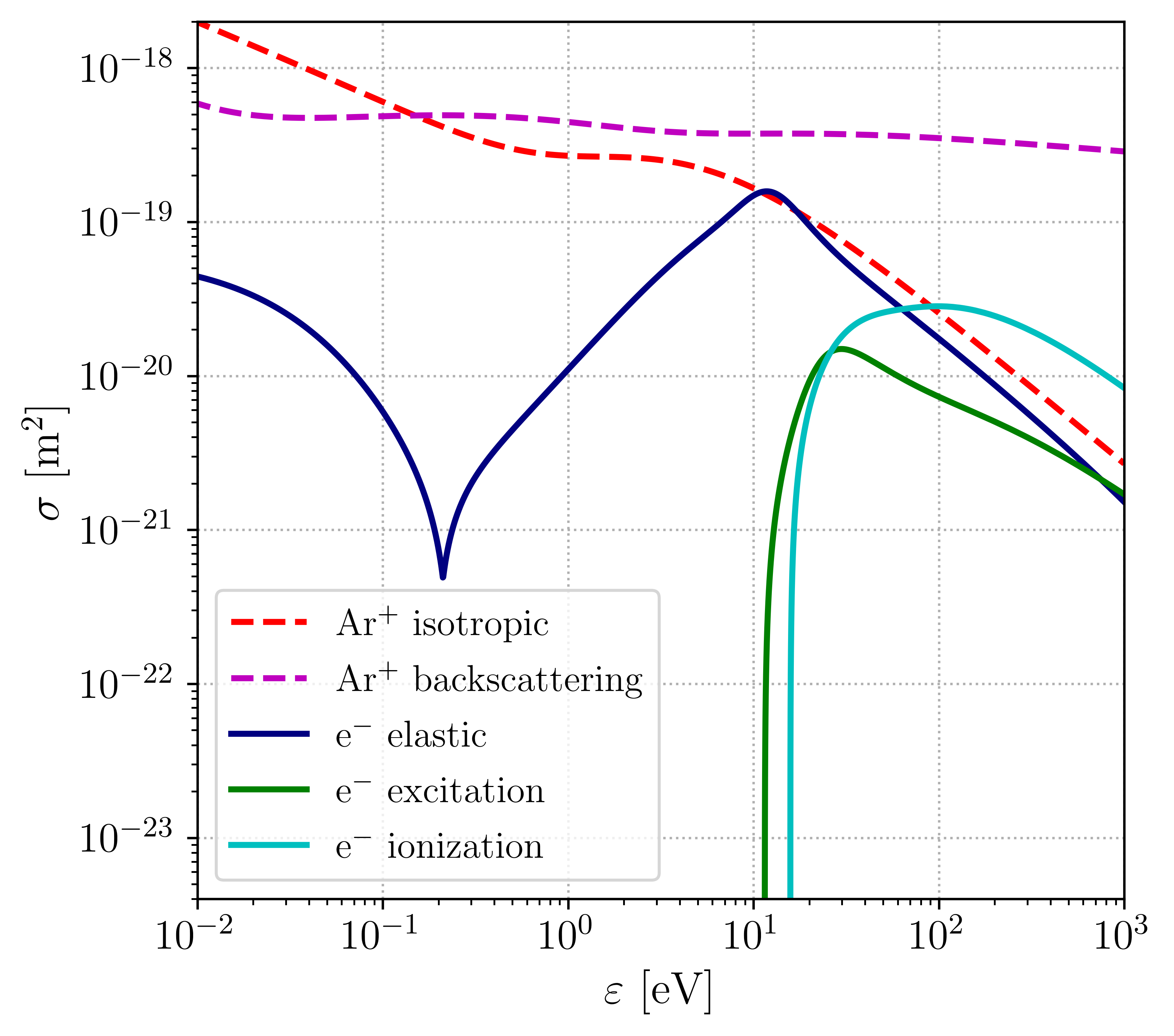}
\caption{Cross sections for the collision processes of electrons (solid lines) and Ar$^+$ ions (dashed lines) considered in the eduPIC code. For the ions, $\varepsilon$ is the center-of-mass energy.}
\label{fig:cross_sections}
\end{center}
\end{figure}

The simulations here are performed in argon gas. The species in the simulation are electrons and Ar$^+$ ions. The electron impact cross sections are adopted from~\cite{phelps1999cold}. This set includes the elastic momentum transfer cross section, one excitation cross section that represents the sum of all excitation cross sections, and the ionization cross section. All e$^{-}+$Ar collisions are assumed to result in isotropic scattering. In the case of Ar$^+$+Ar collisions, only elastic collisions are considered. The cross section set includes an isotropic scattering part, as well as a backward scattering part, as advised in~\cite{phelps1994application} (as already discussed in section \ref{sec:collisions}). 
These cross sections are displayed in figure \ref{fig:cross_sections}.

The cross sections in~\cite{phelps1999cold,phelps1994application} are given by analytic forms. To avoid the need for evaluating these expressions at frequent times when cross section values are needed during the simulation, the functions \texttt{set\_electron\_cross\_sections\_ar()} and \texttt{set\_ion\_cross\_sections\_ar()} tabulate the cross sections at the beginning of the run. The data are stored in the array \texttt{sigma} for electrons and ions, consecutively, with an energy resolution specified by the constant parameter \texttt{DE\_CS} (being equal to 0.001~eV as default).

For the electrons, the array stores the cross section values as a function of the LAB energy, while for the ions, the data are stored as a function of the COM energy. In \cite{phelps1994application}, the data for the ions are given as a function of the laboratory energy, and conversion of the energy is done in the function \texttt{set\_ion\_cross\_sections\_ar()}. The function \texttt{calc\_total\_cross\_sections()} computes the values of $n \sigma_{\rm tot}$, as these are needed for the evaluation of the collision probabilities. These data are stored in the arrays \texttt{sigma\_tot\_e} and \texttt{sigma\_tot\_i}. The constant \texttt{CS\_RANGES} specifies the size of the cross section arrays. (It has to be kept in mind that \texttt{DE\_CS*CS\_RANGES} should exceed the maximum particle energies expected.) During runtime, cross section values are found quickly from these lookup tables.

\subsection{Random number generation} 
\label{sec:rng}

A central component of every Monte Carlo method based simulation is the pseudo-Random Number Generator (RNG). We do not know about a rigorous derivation of the statistical requirements the RNG has to fulfill in order to guarantee the accuracy of the PIC/MCC scheme. Therefore, we rely on our empirical experience and general quality measures. The use of the basic RNG, included in the tested implementations of the C Standard General Utilities Library \texttt{stdlib.h}, was found to result in nonphysical distributions. For this reason, we utilize the Mersenne Twister 19937 generator, included in the C\texttt{++} standard library (beginning with version C\texttt{++}11) in the C and C\texttt{++} versions of the code. The very same RNG is used to generate random samples from (i) the uniform distribution on the [0,1) interval for general purposes, and (ii) the normal distribution used to generate random velocity components of background gas atoms in thermal equilibrium. 

Although the Mersenne Twister RNG is available in Rust in an external library (or ``crate'' as called in Rust terminology), but due to its limited functionality we have chosen to use the \texttt{ThreadRng}, which is a thread-local, cryptographically secure RNG based on the HC-128 algorithm~\cite{Wu2008}.

\subsection{Compilation}

\label{sec:CComp}

These codes have been tested and benchmarked on an Ubuntu Linux computing cluster on nodes equipped with x86-64 based Intel\textsuperscript{\textregistered} Xeon and AMD\textsuperscript{\textregistered} Threadripper CPUs. The compilation of the C and C\texttt{++} codes have been performed with both \texttt{icpc} Intel C\texttt{++} compiler (ver.~2021.1.2, part of the freely available Intel oneAPI Base Toolkit) and \texttt{g++} from the GNU Compiler Collection (ver.~9.3.0). Best performance was achieved on Intel architecture using:\\
\\
\vspace{0.5cm}
\noindent\texttt{icpc -fast -o edupic eduPIC.cc}

\noindent The same compilation line can be used with the C\texttt{++} source file \texttt{eduPIC.cpp} as well. The Rust language compiler is best executed using the integrated package manager \texttt{cargo} with the option \texttt{build}. At default, the compiler creates a ``Debug'' version of the program. To build the final ``Release'' version with high level of compile time optimization turned on, the compile line\\
\\
\vspace{0.5cm}
\noindent\texttt{cargo build -{}-release}\\
can be called from within the project folder.

\subsection{Program execution and files created} 
\label{sec:run}

Before executing the code, the simulation parameters have to be set in the source code, which then has to be compiled as explained above. Subsequently, the code can be invoked in a terminal window as\\ 
\\
\vspace{0.5cm}
\noindent\texttt{./edupic arg1}\\
\noindent To start a new simulation, running an initialization cycle is required by setting \texttt{arg1} to 0, i.e. executing\\ 
\\
\vspace{0.5cm}
\noindent\texttt{./edupic 0}\\
In this case, seeding of a number of initial particles, the simulation of a single RF cycle and saving of the state of the system is executed; further details are given in Algorithm 1 of section \ref{sec:basic_code}.

The code can be run, subsequently, for any number of additional RF cycles specified by \texttt{arg1}, e.g., for 500 cycles, as \\
\\
\vspace{0.5cm}
\noindent\texttt{./edupic 500}\\
\noindent In this run, the previously saved state of the system is restored, the given number of RF cycles is simulated, and at the end of the run, the state of the system is saved. (Please refer to Algorithm 1 of section \ref{sec:basic_code} for more details.) This procedure can be repeated any number of times, the simulation always continues from the previously saved state of the system. During these runs, the time evolution of the number of superparticles is stored in \texttt{conv.dat}, but no other data (results) are saved. 

Measurements on the system can be activated by specifying a second command line argument \texttt{m} when the code is invoked, e.g.,~for 1000 RF cycles, as\\ \\
\vspace{0.5cm}
\noindent\texttt{./edupic 1000 m}\\
\noindent Here all built-in diagnostics are turned on. The number of RF cycles for which measurements are run affects the quality of the statistics of the results. Therefore, we recommend using $\sim$ 1000 RF cycles to obtain results with good signal to noise ratio. When measurements on the system are taken, the code saves the data into the following files:
\begin{itemize}
    \item \texttt{density.dat}: time-averaged density distributions of electrons (second column) and ions (third column) as a function of the position (first column);
    \item \texttt{eepf.dat}: time-averaged EEPF in the central 10\% spatial domain of the discharge (second column) as a function of the energy (first column). The data are normalized corresponding to (the discretized form of) $\int f(\varepsilon) \sqrt{\varepsilon} {\rm d}\varepsilon =1$;
    \item \texttt{ifed.dat}: time-averaged IFEDF at the powered (second column) and grounded (third column)
    electrode as a function of the energy (first column).  The data are normalized corresponding to (the discretized form of) $\int F(\varepsilon) {\rm d}\varepsilon =1$;
    \item \texttt{pot\_xt.dat}: spatio-temporal distribution of the potential;
    \item \texttt{efield\_xt.dat}: spatio-temporal distribution of the electric field;
    \item \texttt{ne\_xt.dat}: spatio-temporal distribution of the electron density;
    \item \texttt{ni\_xt.dat}: spatio-temporal distribution of the ion density;    
    \item \texttt{je\_xt.dat}: spatio-temporal distribution of the electron current density;
    \item \texttt{ji\_xt.dat}: spatio-temporal distribution of the ion current density;
    \item \texttt{powere\_xt.dat}: spatio-temporal distribution of the power absorption by the electrons;
    \item \texttt{poweri\_xt.dat}: spatio-temporal distribution of the power absorption by the ions;
    \item \texttt{meanee\_xt.dat}: spatio-temporal distribution of the mean electron energy;
    \item \texttt{meanei\_xt.dat}: spatio-temporal distribution of the mean ion energy;
    \item \texttt{ioniz\_xt.dat}: spatio-temporal distribution of the ionization rate.
\end{itemize}

The number of the energy bins as well as the resolution of the EEPF and IFEDF are set by the constants \texttt{N\_EEPF}, \texttt{DE\_EEPF} and \texttt{N\_IFED}, \texttt{DE\_IFED}.

The ``xt files'' that store the spatio-temporal variation of the given quantity contain a specific number of rows related to the number of grid cells \texttt{N\_G}. The number of columns represents the temporal resolution. In order to improve the signal to noise ratio of the data, the number of time steps per RF cycle \texttt{N\_T} is binned to a lower number \texttt{N\_XT}. The number of time steps binned for the xt files can be set by the variable \texttt{N\_BIN} in the code. (At this point, care should be taken to ensure that \texttt{N\_T} is an integer multiple of \texttt{N\_BIN}.)
All the output data are in SI units, except for the mean electron energy, as well as for the EEPF and the IFEDF, for which eV units are used. 

Following a run in the ``measurement'' mode, the code saves a file named \texttt{info.txt}. This file contains the operation parameters and simulation settings as a record of the run. The file also displays information about the stability and accuracy conditions. First, the conditions 1--3 (see section \ref{sec:stability}) are checked. These concern the relation of the grid spacing to the Debye length, the relation of the time step to the electron plasma frequency, and the collision probabilities during a time step. Whenever any of these checks fail, an error message is issued and no further diagnostics data are saved. 

We note that the evaluation of the stability and accuracy criteria is meaningful only when measurements are taken over the converged state. The initial setting of the simulation parameters (like the grid size and the time steps) is normally based on an educated guess, these parameters can be refined later according to the information contained in \texttt{info.txt}. In case of need of modifications of the parameter settings, the code has to be re-compiled, the simulation has to be re-converged, and the stability and accuracy criteria have to be checked again.

In case the above conditions are met, the diagnostics data listed above are saved to the respective files, and the maximum electron energy for which the Courant–Friedrichs–Lewy condition (condition 4 in the list of section \ref{sec:stability}) still holds
at the actual values of \texttt{N\_G} and \texttt{N\_T}, is also displayed in \texttt{info.txt}. To make sure that this condition holds for the {\it vast majority} of the electrons, one has to confirm that the EEPF decays to a small value at this energy. This is not done automatically in the code, the procedure is left to the user. We advise to observe the EEPF obtained in the center of the plasma, and to use a threshold value of $f(\varepsilon) \sim 10^{-6}$ eV$^{-3/2}$. For a more rigorous check, the complete space- and time-resolved EEPF should be considered.

Additional information about the particle characteristics at the electrodes and about the power absorption by the electrons and ions is also saved to \texttt{info.txt} at the end of the simulation. The latter is computed as the spatial and temporal average of the product $j(x,t) E(x,t)$. This product is also saved to the files \texttt{powere\_xt.dat} and \texttt{poweri\_xt.dat} (corresponding to the power absorption rate by the electrons and ions, respectively) with spatial and temporal resolution.

\subsection{Basic (C) code}
\label{sec:basic_code}

Below, a detailed explanation of the structure and the operation of the C code is given at the level of algorithms. Algorithm \ref{alg:mainfunction} presents the \texttt{main()} function of the simulation code. Subsequently, in algorithms \ref{alg:charge}--\ref{alg:collisions}, details of the six steps of the PIC/MCC cycle (see section 2.1) are presented, which are  parts of the \texttt{do\_one\_cycle} function. In algorithms \ref{alg:charge}-\ref{alg:boundaries}, the calculation routines are only shown for the electrons. Similar steps are applied for the positive ions, however, these are only performed every N\_SUB-th time step as ions move slower in the plasma. By this subcycling method, remarkable computation time is gained without decreasing the accuracy of the calculation. In algorithm \ref{alg:collisions}, though, which shows the collision routine, the steps applied for ions are also shown, since they are considerably  different from the ones applied for electrons.

Measurements of the relevant physical quantities (listed above) are carried out at two points within the \texttt{do\_one\_cycle} function. Accumulation of the data for the quantities that are defined at the grid points (potential, electric field, electron and ion density) is performed following the six steps of the PIC/MCC cycle. Data for some other quantities is accumulated in the part of the code where the particles are moved. The reason for this is that the positions and the velocities of the particles are not defined at the same time (recall that positions are defined at integer time steps, while position at half-integer time steps, see section \ref{sec:six_steps}.). For an accurate calculation of, e.g., the power absorption by the particles their velocities have to be known at integer time steps as well. This is solved in the code by computing the particle velocities at an intermediate time as well (i.e., at the same time when the positions are known) and the measured quantities (e.g., mean velocity, mean energy) are evaluated with these average velocities. The measured quantities are updated at cells of two-dimensional data structures (``xt arrays'') with a similar interpolation like the one used at the density assignment of the particles to the grid points.
The details of these calculations are omitted from the forthcoming discussion of the algorithms of the PIC/MCC cycle.

\begin{algorithm}
\label{alg:mainfunction}
	\DontPrintSemicolon
	
	\vspace{0.5cm}

    \tcp{get an integer parameter from command line that controls code execution and number of RF cycles to be simulated:}
    
    strcpy(st0, argv[1])
    
    arg1 = atol(st0)
	
	\tcp{activate measurement mode if m is given as the second command line argument:}
	\uIf {\rm{argv[2] = m}}
	{
		measurement\_mode = true
	} 
	\Else {
		measurement\_mode = false
	}
	
	\tcp{call functions to compute and tabulate the cross sections:}
	
	set\_electron\_cross\_sections\_ar()
	
	set\_ion\_cross\_sections\_ar()
	
	calc\_total\_cross\_sections()
	
	\tcp{perform the simulation of a given number of RF cycles:}
	\uIf {\rm{arg1 = 0}} 
	{
		\tcp{perform an initialization cycle:}
		no\_of\_cycles = 1 
		
		init(N\_INIT) \tcp*{seed initial particles}
		
		Time = 0 \tcp*{initialize simulation time}
		
		do\_one\_cycle() \tcp{simulate one RF cycle}
		
		cycles\_done = 1
	} 
	\Else 
	{
		no\_of\_cycles = arg1 \tcp*{number of RF cycles to simulate}
		
		load\_particle\_data()  \tcp*{load the state of the system}
		                           
		\tcp{simulate the given number of RF cycles:}
		\For{\rm{0 $\leq$ i < no\_of\_cycles}}
		{
			do\_one\_cycle()
		}
	
		cycles\_done += no\_of\_cycles
	}
	
	save\_particle\_data() \tcp*{save the state of the system}
	
	\tcp{save results when measurements are activated:}
	\If {\rm{measurement\_mode}}
	{
		save\_density()
		
		save\_eepf()
		
		save\_ifed()
		
		save\_all\_xt()
		
		check\_and\_save()
	}
	
	\vspace{0.5cm}
	\caption{The \texttt{main()} function of the program. When the code is started with a command line argument \texttt{arg1}=\texttt{0}, a given number (\texttt{N\_INIT}) of electrons and Ar$^+$ ions are seeded within the gap at random positions and with zero initial velocities. Then, the simulation of a single RF cycle is executed by the \texttt{do\_one\_cycle()} function. Following this, the number of the particles and their coordinates are saved to the \texttt{picdata.bin} binary file. When the code is invoked with an integer command line argument \texttt{arg1}$>0$, the code simulates further \texttt{arg1} RF cycles. This run starts with loading the state of the system from the file \texttt{picdata.bin}. Then, \texttt{do\_one\_cycle()} is called  \texttt{arg1} times, and finally the state of the system is saved again to \texttt{picdata.bin}. During these runs, the time evolution of the number of superparticles is stored in the file \texttt{conv.dat}, but no additional measurements on the system are being conducted. When a second command line argument \texttt{m} is specified, measurements on the system are activated and the data acquired are saved to data files (as described above).} 
\end{algorithm}

\begin{algorithm} \label{alg:charge}
	\DontPrintSemicolon
	\vspace{0.5cm}
	$\rm{FACTOR\_W = WEIGHT / DV}$ \tcp*{scaling factor for the density} 
	
	\For{$0\leq$\rm{p}$<$\rm{N\_G}}
	{
		$\rm{e\_density[p] = 0}$
	}
	
	\For{$0\leq$\rm{k}$<$\rm{N\_e}}
	{	
	$\rm{c0 = x\_e[k] * INV\_DX }$
	
		$\rm{p = (int)(c0)}$
		\tcp*{grid index} 
		
		$\rm{ e\_density[p]   += (p + 1 - c0) * FACTOR\_W }$
		
		$\rm{ e\_density[p+1] += (c0 - p) * FACTOR\_W }$
		}
	
	$\rm{e\_density[0]*= 2.0 }$
	
	$\rm{e\_density[N\_G-1] *= 2.0}$
	\vspace{0.5cm}	
	\caption{PIC/MCC step 1 -- Computation of the charged particle densities. At the beginning of each RF cycle, the electron density is set to zero at the grid points. Then, for each of the electrons, the density at the two grid points which enclose the given electron, is incremented. Finally the electron density at the two edge points of the grid is multiplied by two, because the volume that belongs to these points is the half of that compared to the other grid points.}
\end{algorithm}

\begin{algorithm} \label{alg:Poisson}
	\DontPrintSemicolon
	\vspace{0.5cm}		
	\For{$0\leq$\rm{p}$<$\rm{N\_G}}
	{
		$\rm{rho[p] = E\_CHARGE * (i\_density[p] - e\_density[p]) }$ 
	}
	solve\_Poisson(rho,Time)
	
	\tcp{calculation of the potential and the electric field at the grid points}

	\vspace{0.5cm}
	\caption{PIC/MCC step 2 -- Computation of the potential and the electric field. The discretized Poisson equation can be written in a matrix form with a tridiagonal matrix of constant coefficients. The equation is solved using the Thomas algorithm~\cite{Thomas1949}. The electric field values at the grid points are also derived within the \texttt{solve\_Poisson()} function. }
\end{algorithm}

\begin{algorithm}\label{alg:motion}
	\DontPrintSemicolon
	\vspace{0.5cm}	
	
	FACTOR\_E = DT\_E / E\_MASS * E\_CHARGE 
	\tcp*{scaling factor for the change of velocity}

	\For{$0\leq$\rm{k}$<$\rm{N\_e}}
	{
	    c0 = x\_e[k] * INV\_DX
	    
		p = (int)(c0) \tcp*{grid index}
		c1  = p + 1 $-$ c0 \tcp*{linear interpolation factors}
        
        c2  = c0 $-$ p
        
        e\_x = c1 $*$ efield[p] + c2 $*$ efield[p+1] 
		
		$\rm{vx\_e[k] -= e\_x * FACTOR\_E }$
		
		$\rm{x\_e[k]  += vx\_e[k] * DT\_E}$
	}
	
	\vspace{0.5cm}
	\caption{PIC/MCC steps 3 and 4 -- Computation of the forces acting on the particles and moving the particles. For each electron, first, the velocity is updated according to the electric field interpolated to its position from the values at the grid points that enclose the given electron. Subsequently, the position is updated according to the leapfrog scheme.}
\end{algorithm}

\begin{algorithm}\label{alg:boundaries}
	\DontPrintSemicolon
	\vspace{0.5cm}	
	
	k = 0

	\While{$\rm{k < N\_e}$}
	{
	    \tcp{cycle over all electrons:}
		out = false 

		\If {$\rm{x\_e[k] < 0}$ } 
		{
			N\_e\_abs\_pow++  
			
			out = true 
		}
		
		\If {$\rm{x\_e[k] > L}$ } 
		{
			N\_e\_abs\_gnd++  
			
			out = true  
		}
		
		\uIf {$\rm{out}$} 
		{		
			\tcp{remove electron:}
			
			$\rm{x\_e [k] = x\_e [N\_e-1]}$
			
			$\rm{vx\_e[k] = vx\_e[N\_e-1]}$
			
			$\rm{vy\_e[k] = vy\_e[N\_e-1]}$
			
			$\rm{vz\_e[k] = vz\_e[N\_e-1]}$
			
			$\rm{N\_e--}$
		} 	
		\Else{ 
		k++
		}
}
\vspace{0.5cm}
\caption{PIC/MCC step 5 -- Adding/removing particles at the boundaries. Electrons whose updated position is outside the electrode gap are identified and removed. In such cases, the boolean variable \texttt{out} (originally set to false) is set to true, the counter of the number of absorbed electrons at the given electrode is incremented, and the electron is removed from the array that stores the electron coordinates. Finally, \texttt{N\_e} is decremented. For the ions (not shown here), the energy of the ion upon its impact at the electrode is used for building the IFEDF. Recall that we do not consider the reflection and emission of particles at/from the electrode surfaces.}
\end{algorithm}

\begin{algorithm} \label{alg:collisions}
	\DontPrintSemicolon
	\vspace{0.5cm}
	
	\For {$\rm{0 \leq k < N\_e}$}
	{                              
		\tcp{calculate the kinetic energy of the electron:}
		
		v\_sqr = vx\_e[k] * vx\_e[k] + vy\_e[k] * vy\_e[k] + vz\_e[k] * vz\_e[k]
		
		velocity = sqrt(v\_sqr)
		
		energy = $\rm{0.5 * E\_MASS * v\_sqr / EV\_TO\_J}$
		
		\tcp{obtain the collision frequency of the electron with the background gas:}
		
		energy\_index = min( int(energy / DE\_CS + 0.5), $\rm{CS\_RANGES-1)}$ \tcp*{variable indexing the energy bin the electron belongs to}
		
		nu = sigma\_tot\_e[energy\_index] * velocity
		
		\tcp{calculate the collision probability for the electron:}
		
		$\rm{p\_coll = 1 - \exp(- nu * DT\_E)}$

		\tcp{perform electron--atom collision with the probability obtained, and increase the collision counter:}
		
		\If {$\rm{R01(MTgen) < p\_coll}$} 
		{                      
			collision\_electron(x\_e[k], vx\_e[k], vy\_e[k], vz\_e[k], energy\_index)
			
			N\_e\_coll++
		}
	}
	
	\If {$\rm{(t ~mod~ N\_SUB) = 0}$} 
	{
		\For {$\rm{0 \leq k < N\_i}$}
		{
			\tcp{draw velocity components of a random target gas atom from the Maxwell-Boltzmann distribution:}
			
			vx\_a = RMB(MTgen)                          
			
			vy\_a = RMB(MTgen)
			
			vz\_a = RMB(MTgen)
			
			\tcp{calculate the relative velocity of the collision partners and the kinetic energy of the projectile:}
			
			$\rm{gx  = vx\_i[k] - vx\_a }$
			
			$\rm{gy  = vy\_i[k] - vy\_a }$
			
			$\rm{gz  = vz\_i[k] - vz\_a }$
			
			$\rm{g\_sqr = gx * gx + gy * gy + gz * gz }$
			
			$\rm{g = sqrt(g\_sqr) }$
			
			$\rm{energy = 0.5 * MU\_ARAR * g\_sqr / EV\_TO\_J }$
			
			\tcp{obtain the collision frequency of the ion with the background gas:}
			
			energy\_index = min( int(energy / DE\_CS + 0.5), $\rm{CS\_RANGES-1)}$ \tcp*{variable indexing the energy bin the ion belongs to}
			
			nu = sigma\_tot\_i[energy\_index] * g
			
			\tcp{calculate the collision probability for ions:}
			
			$\rm{p\_coll = 1 - \exp(- nu *  DT\_I)}$
			
			\tcp{perform ion--atom collision with the probability obtained, and increase the collision counter:}
			
			\If {$\rm{R01(MTgen) < p\_coll}$} 
			{                   
				collision\_ion (vx\_i[k], vy\_i[k], vz\_i[k], vx\_a, vy\_a, vz\_a, energy\_index)
				
				N\_i\_coll++
			}
		}
	}
	
	\vspace{0.5cm}
	\caption{PIC/MCC step 6 -- Checking and executing collisions. First, the collisions of the electrons are considered. Subsequently,  the collisions of ions are handled, however, only in every \texttt{N\_SUB}-th time steps, in accordance with ion subcycling. (\texttt{t} is the loop variable for the time steps within an RF cycle.) For the ions, the potential collision partners are sampled from the thermal distribution of the background gas atoms.}
\end{algorithm}

\subsection{C\texttt{++} code} \label{sec:cpp_code}

The C programming language falls into the category of ``procedural imperative'' languages, in which the programmer instructs the machine how to change its state and instructions are grouped into procedures~\cite{Programming}. Representing one of the simplest programming paradigms~\cite{Paradigm} it is best suited for short codes and in cases where tight control over the hardware resources and timing is necessary. For a long time, its low level ``close to metal'' nature made it popular (as an alternative to Fortran) in high performance computing (HPC) applications, where the codes are optimized to perform one particular computational task.

With the introduction of the C\texttt{++} programming language, another level of abstraction was added realizing the ``object oriented'' programming paradigm, which groups instructions with the part of the state they operate on. This pawed the way to the construction of large, multi-developer and multiple purpose programs, like modern operating systems and office program suites. Clearly, for the purposes of this study and the small size of the eduPIC code the programming overhead of object oriented implementation would not pay off, but it is a logical direction for further development targeting, e.g., a more complicated gas composition and chemistry. Another component of modern C\texttt{++}, as maintained by the ``Standard C\texttt{++} Foundation'' is the ``standard library'' (\texttt{std}), a large collection of functions, types and objects of which the functionality is strictly defined but the implementations do continuously evolve and are optimized to a variety of hardware architectures.

The C\texttt{++} implementation of the eduPIC code closely follows the structure of the C version but utilizes modern data structures, like \texttt{std::array}-s and \texttt{std::vector}-s, as well as a variety of vector transformation standard library routines. To illustrate the difference, let us focus on the first \texttt{for} loop in Algorithm \ref{alg:Poisson}, the simple calculation of the charge density from the particle density distributions: in the C code, (i) a temporary variable \texttt{p} had to be declared and initialized, (ii) the limit and increment of the loop had to be set, (iii) the input values had to be fetched from memory addresses calculated from the head-of-array addresses and the offsets given by \texttt{p}, (iv) the result of the calculation had to be directed to a specific memory address inside the output array. All these elementary steps had to be defined in the program explicitly. The disadvantage of having to define so many details is twofold: (i) errors can be introduced at any of these steps, (ii) by defining each elementary step and with this the order of execution, there remains very limited room for the compiler to perform optimization to the target hardware architecture. This latter point becomes increasingly important taking into account that the development of computing hardware in the last decade increasingly utilized multiple levels of parallelism (e.g. vector registers, multiple cores in CPUs, multiple CPUs per node, as well as massively parallel GPU acceleration options) at an increasing pace. Replacing the \texttt{for} loop by the \texttt{std::transform} function improves on both points, less details have to be defined and the best performing machine level code can be chosen by the compiler on the actual hardware architecture, including optional asynchronous parallel versions, if proper execution policies are applied. Similar statements can be formed about many other parts of the code, which are, however, not discussed further here. 

\subsection{Rust code} \label{sec:rust_code}

The Rust-language is a modern open-source, community-developed programming language maintained by the ``Rust Core Team'' with its first stable release published in 2015. Rust is a multi-paradigm programming language, syntactically similar to C\texttt{++}, designed for performance and memory/thread safety~\cite{RustBook}. The advantage of using a new but mature programming language is that one can utilize all successful modern tools without the bounds set by backward compatibility requirements to already superseded old concepts. Similarly to C\texttt{++}, all modern data structures like dynamic vectors, as well as a rich standard library are available. The Rust compiler (\texttt{rustc}) and package manager (\texttt{cargo}) provide user friendly error messages and effortless download, integration and version control of code libraries. The default behavior of the Rust language elements are tuned for best performance and safety. As an example, variables by default are immutable objects, meaning that after declaration their value can be initialized once during run-time and are read-only from there on within their scope (in contrast to constants, where the values have to be known already at compile-time). This introduces obvious optimization possibilities to the compiler and during execution, like cache memory management. Regarding the safety concerns, the handling of arrays and associated pointers are strict, making it virtually impossible to accidentally read or write out of bounds and trigger segmentation faults, or even worse, undefined behavior of the program.

Despite performing the same computations as the base C version, the Rust implementation of the eduPIC simulation features some adaptations in order to exploit certain features of the language. Particle data (position and velocity components) are stored in a vector-of-structures dynamic container making particle management (adding and removing particles) easy and safe in contrast to the base code, where no intrinsic mechanism guarantees the consistency of the particle counters and the length of the independent position and velocity component arrays.  Syntax wise a very appealing feature is that arithmetic function calls can be applied in an object-oriented style, which improves readability of complex expressions. Although in such a short and single source file code global variables do not represent safety hazards, following the general concept mostly relevant to more complex projects, we have omitted global variables from the Rust code.

The execution of the Rust version of the simulation can be performed by calling the executable file as described in section \ref{sec:run}, or using the package manager \texttt{cargo} as \\
\\
\vspace{0.5cm}
\noindent\texttt{cargo run -{}-release 1000 m}\\
\noindent In this case the highly optimized ``Release'' version with arguments \texttt{1000} (number of RF cycles to simulate) and \texttt{m} (measurements turned on) will be executed after additional package download and code compilation steps, if needed to build the executable file, are performed.

\vspace{1cm}

The three implementations (C, C\texttt{++} and Rust) described above, each utilizing different levels of abstractions and modern programming language features, are available to the interested reader \cite{eduPIC_SOURCEs}. All three implementations provide comparable execution times and equivalent simulation results of the computed physical quantities.

\section{Results}

In this section, first we present some ``reference results'' that were obtained with the eduPIC code  with a set of ``default'' discharge parameters (section \ref{sec:r1}). These results are useful for demonstrating the capabilities of the code, as well as the basic operation characteristics of the simple CCP considered here. Subsequently, we address a physics problem, which can readily be studied without any extension of the code: in section \ref{sec:r3} we illustrate the formation of the flux-energy distribution of Ar$^+$ ions at the electrode surfaces for the conditions of dual-frequency excitation. We note, that for all these calculations the number of superparticles was around 10$^5$, which normally provides acceptable accuracy, however, higher numbers are recommended for more accurate results \cite{turner2006kinetic,Erden}.


\subsection{Reference results}
\label{sec:r1}

Below, we illustrate the reference results obtained with our code. The discharge is driven by a single harmonic voltage waveform:
\begin{equation}
    V(t) = V_0 \cos (2 \pi f t).
\end{equation}
$V_0$ and $f$, along with the other simulation parameters are listed in table~\ref{tab:table2}. When downloaded, compiled, and executed, the code should reproduce these results with its ``default'' parameters included in the source files. The following figures have been generated with Python and the corresponding plotting library Matplotlib. 

\begin{table} [ht]
\centering%
\caption{Set of ``default'' parameters for the reference simulation run.}
\begin{tabular}{ll}
\hline
Quantity & Value \\ 
\hline
Driving voltage amplitude & $V_0$ = 250 V \\
Driving frequency & $f$ = 13.56 MHz \\
Superparticle weight & $W$ = 7$\times10^4$\\
Electrode gap & $L$ = 25 mm \\
Ar pressure & $p$ = 10 Pa \\
Gas temperature & $T_{\rm g}$ = 350 K \\
Number of grid points & $N_{\rm g}$ = 400 \\
Number of time steps / RF cycle & $N_{\rm t}$ = 4000 \\
\hline
\end{tabular}
\label{tab:table2}
\end{table}

\begin{figure}[h!]
\centering
\begin{center}
\includegraphics[width=0.4\textwidth]{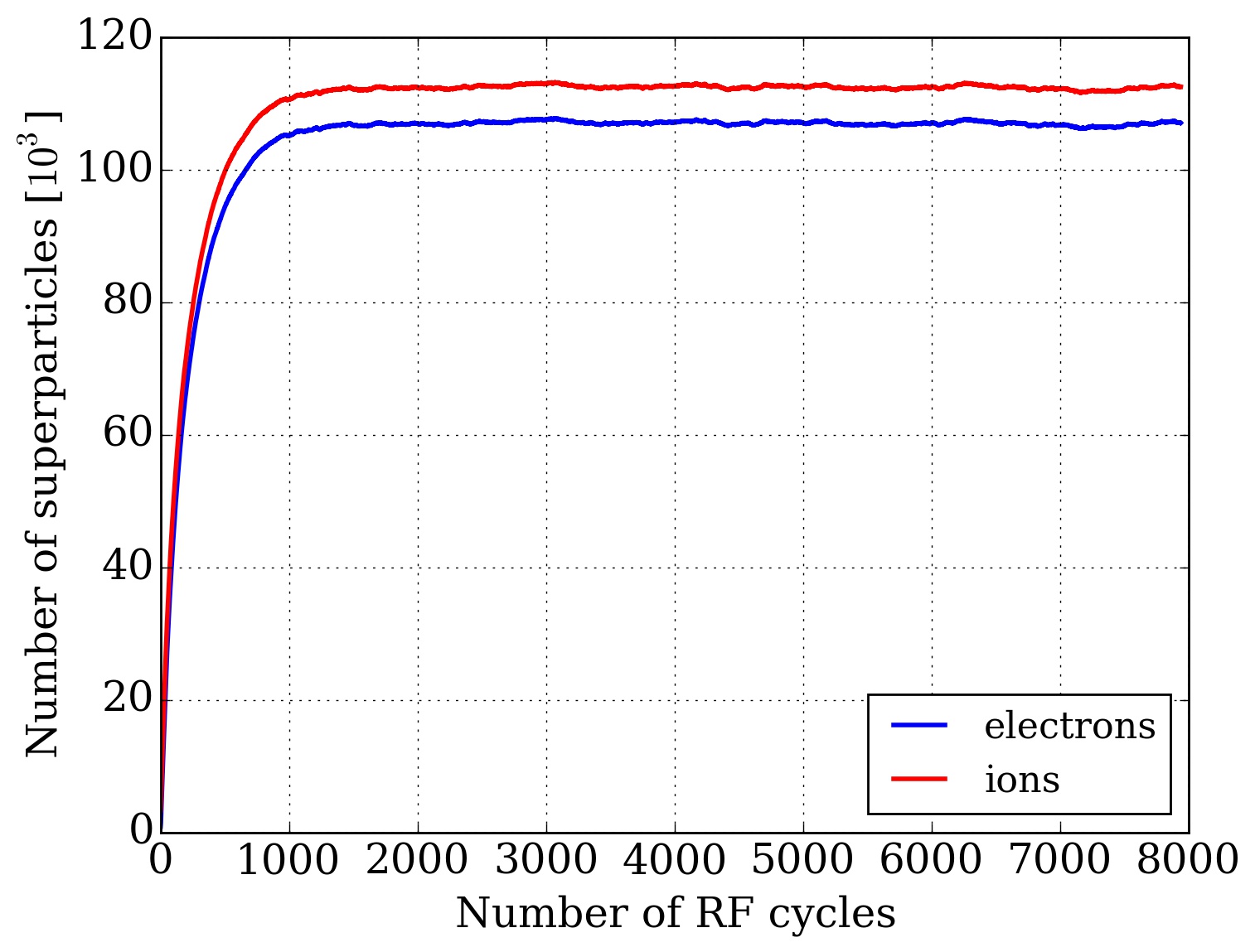} 
\caption{The number of superparticles as a function of the number of RF cycles (file: \texttt{conv.dat}). Discharge conditions: argon $p=10$ Pa, $L=25$ mm, $T_{\rm{g}} = 350$ K, $V_{\rm{}} = 250$~V and $f_{\rm{}} = 13.56$ MHz.}
\label{fig:convergence}
\end{center}
\end{figure}

\begin{figure}[h!]
\centering
\begin{center}
\includegraphics[width=0.4\textwidth]{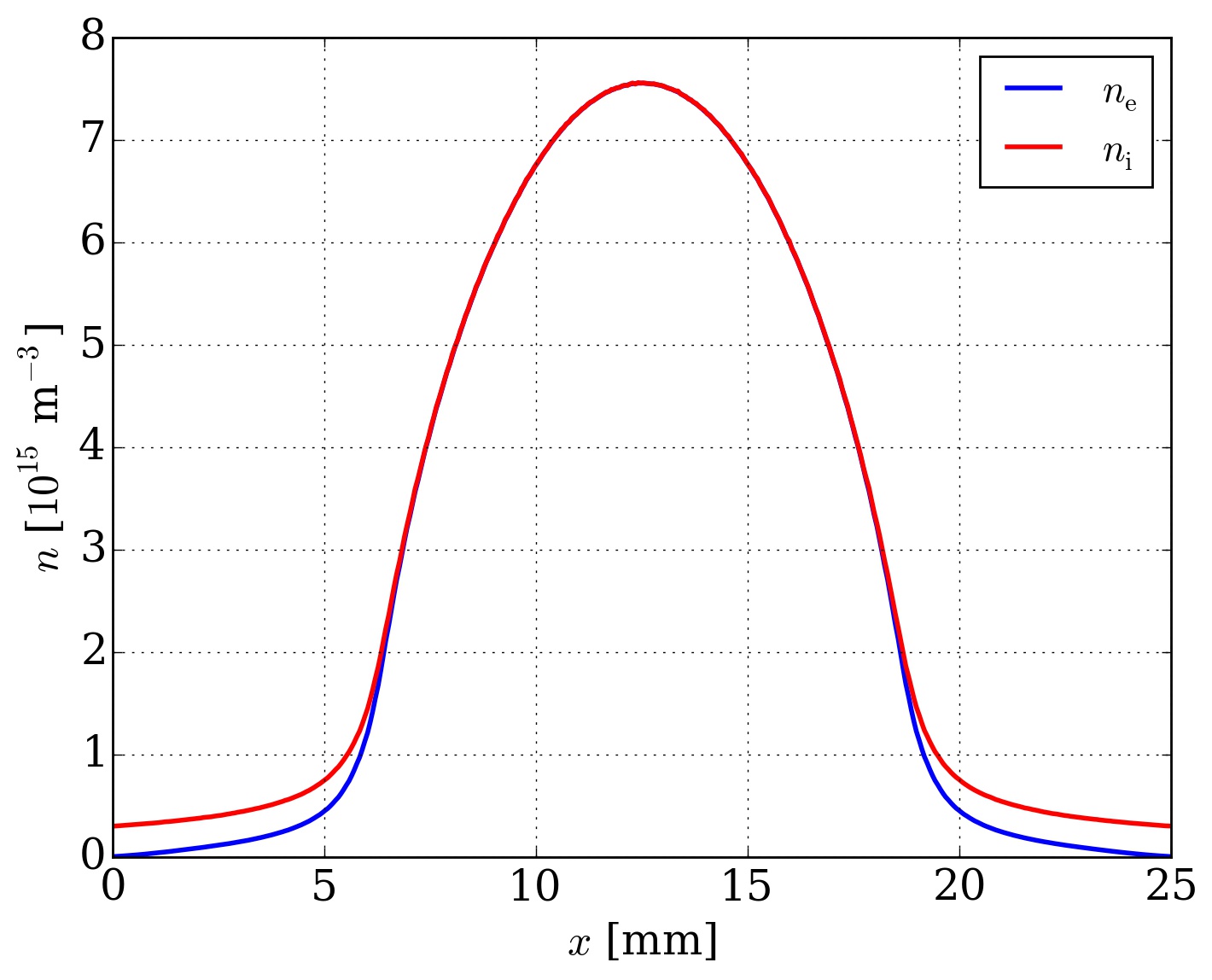} 
\caption{The temporally averaged electron (blue) and ion density (red) as a function of the spatial position between the electrodes (data file: \texttt{density.dat}). Discharge conditions: argon $p=10$ Pa, $L=25$ mm, $T_{\rm{g}} = 350$~K, $V_{\rm{}} = 250$ V and $f_{\rm{}} = 13.56$ MHz.}
\label{fig:density}
\end{center}
\end{figure}

Figure \ref{fig:convergence} shows the dependence of the number of superparticles in the simulation on the RF cycles elapsed after the initialization of the simulation with 1000 initial superparticles of both species. One can see that convergence is reached after $\sim$1500 cycles. At this stage, the number of superparticles as a function RF cycles (or time) does not change anymore, it only fluctuates around the equilibrium value. During this initial phase no data for the plasma characteristics were collected except for the number of superparticles. An additional run for 1000 cycles was executed to collect the data shown in the subsequent figures. Recall, that the code does not consider any surface processes. 

Figure \ref{fig:density} shows the temporally averaged electron and ion density as a function of space. This plot clearly illustrates the formation of a quasi-neutral plasma bulk in the center of the discharge and the two electron depletion regions close to both electrodes which are defined as the plasma sheaths.

\begin{figure*}[h!]
\centering
\begin{center}
\includegraphics[width=0.95\textwidth]{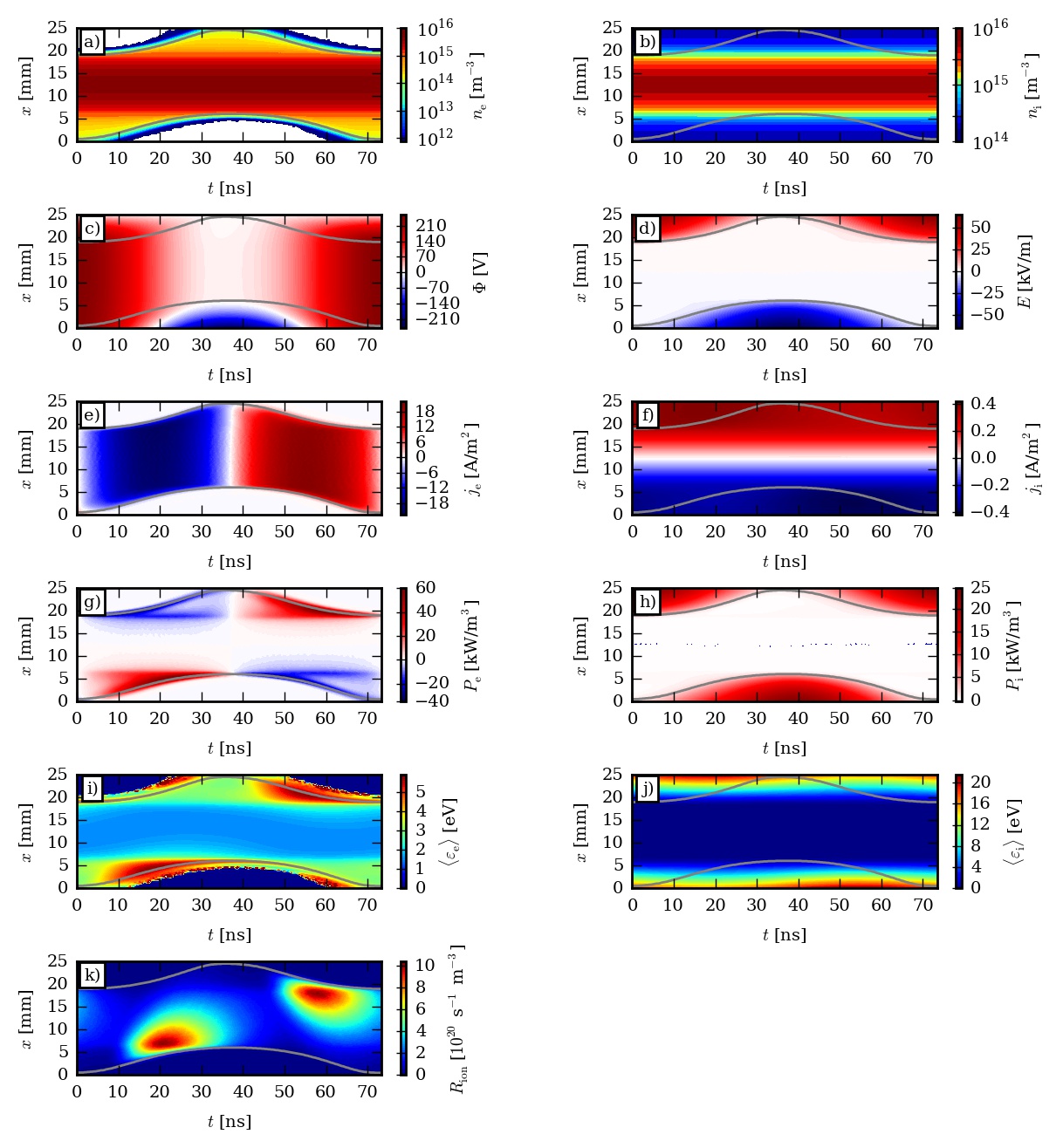} 
\caption{Spatially and temporally resolved distributions of (a) the electron density, $n_{\rm e}$, in m$^{-3}$ (data file: \texttt{ne\_xt.dat}), (b) the ion density, $n_{\rm i}$, in m$^{-3}$ (data file: \texttt{ni\_xt.dat}), (c) the potential, $\Phi$, in V (data file: \texttt{pot\_xt.dat}), (d) the electric field, $E$, in kV/m (data file: \texttt{efield\_xt.dat}), (e) the electron current density, in A/m$^2$ (data file: \texttt{je\_xt.dat}), (f) the ion current density, in A/m$^2$ (data file: \texttt{ji\_xt.dat}), (g) the electron power density, in kW/m$^3$ (data file: \texttt{powere\_xt.dat}), (h) the ion power density, in kW/m$^3$ (data file: \texttt{poweri\_xt.dat}), (i) the mean electron energy, $\langle\varepsilon_{\rm{e}}\rangle$, in eV (data file: \texttt{meanee\_xt.dat}), (j) the mean ion energy, $\langle\varepsilon_{\rm{i}}\rangle$, in eV (data file: \texttt{meanei\_xt.dat}) and (k) the ionization rate, $R_{\rm ion}$, in s$^{-1}$m$^{-3}$ (data file: \texttt{ioniz\_xt.dat}). The  horizontal axis shows the time for one RF cycle and the vertical axis shows the distance from the powered electrode. The gray solid lines indicate the plasma sheath edge, calculated using the formula given in \cite{brinkmann2007beyond}. Discharge conditions: argon $p=10$ Pa, $L=25$ mm, $T_{\rm{g}} = 350$ K, $V_{\rm{}} = 250$ V and $f_{\rm{}} = 13.56$ MHz.}
\label{fig:XT10Pa}
\end{center}
\end{figure*}

In figure \ref{fig:XT10Pa}, we display the spatio-temporal distributions of several discharge characteristics: the density of the electrons and the ions (panels (a), (b)), the potential and the electric field (panels (c), (d)), the electron and ion current density (panels (e), (f)), the power density absorbed by the electrons and the ions (panels (g), (h)),
the mean energy of the electrons and ions (panels (i), (j)), as well as  and the ionization source function (panel (k)). 

Panels (a) and (b) clearly reveal the marked difference between the temporal dynamics of the electrons and the positive ions. While the spatial distribution of the ions is nearly time-independent, the motion of the electrons is governed by the temporal variation of the potential and the electric field (shown in panels (c) and (d), respectively). The electron density is depleted within the sheath regions that form near both electrodes with a shift of $T_{\rm RF}/2$ (where $T_{\rm RF}=1/f$). The edge of the sheaths, found as advised in \cite{brinkmann2007beyond}, is indicated by the gray lines in each panel. Figures \ref{fig:XT10Pa}(e) and (f) show the electron and ion current density, respectively. The electron current density is dominant in the plasma bulk (absolute values around 20 A/m$^2$) and almost zero within the sheath regions, due to electron depletion (see figure \ref{fig:XT10Pa}(a)). The ion current density increases from the centre of the discharge towards the electrodes from 0 to $\pm0.4$ A/m$^2$. The continuity of the total current is ensured by a significant displacement current (not shown) within the sheath regions.

The power absorption of the electrons (figure \ref{fig:XT10Pa}(g)) is prominent near the edges of the expanding sheaths. Upon sheath collapse, this quantity acquires negative values, corresponding to power loss. The ``vertical'' features observable in this plot near the positions of the maximum extent of the sheaths show the effect of the ambipolar electric field (see e.g., \cite{heat1}). The positive ions absorb power mostly within the sheath regions while they move towards the electrodes under the influence of the high electric field present in this domain (figure \ref{fig:XT10Pa}(h)). For these parameter settings the space- and time-averaged power density absorbed by the electrons is  $1.58 \times 10^3$ W\,m$^{-3}$, while that absorbed by the ions is about a factor of two higher, $3.28 \times 10^3$ W\,m$^{-3}$.

As a consequence of the power absorption by the electrons in the vicinity of the edges of the expanding sheaths the mean energy of the electrons also exhibits a maximum in this domain of space and time (see figure \ref{fig:XT10Pa}(i)). Additional, smaller maxima of the mean energy are also observed in the vicinity of the electrodes, at times when the sheaths collapse; these are caused by electric field reversals \cite{schulze2008electric}. The mean energy of the electrons within the bulk plasma is nearly constant, around 1.5--2 eV. The mean energy of the ions, as illustrated in figure \ref{fig:XT10Pa}(j), is only slightly modulated in time and reaches maximum values of about 20 eV near both electrodes. The power absorption and the enhanced mean electron energy in the vicinity of the edges of the expanding sheaths \cite{heat1,heat2,heat3}, gives rise to a significant ionization (see figure \ref{fig:XT10Pa}(k)) within these times of the RF period, however, over a more extended spatial scale, as compared figure \ref{fig:XT10Pa}(g), due to the acceleration that the electrons experience towards the center of the discharge. 

\begin{figure}[h!]
\centering
\begin{center}
\includegraphics[width=0.4\textwidth]{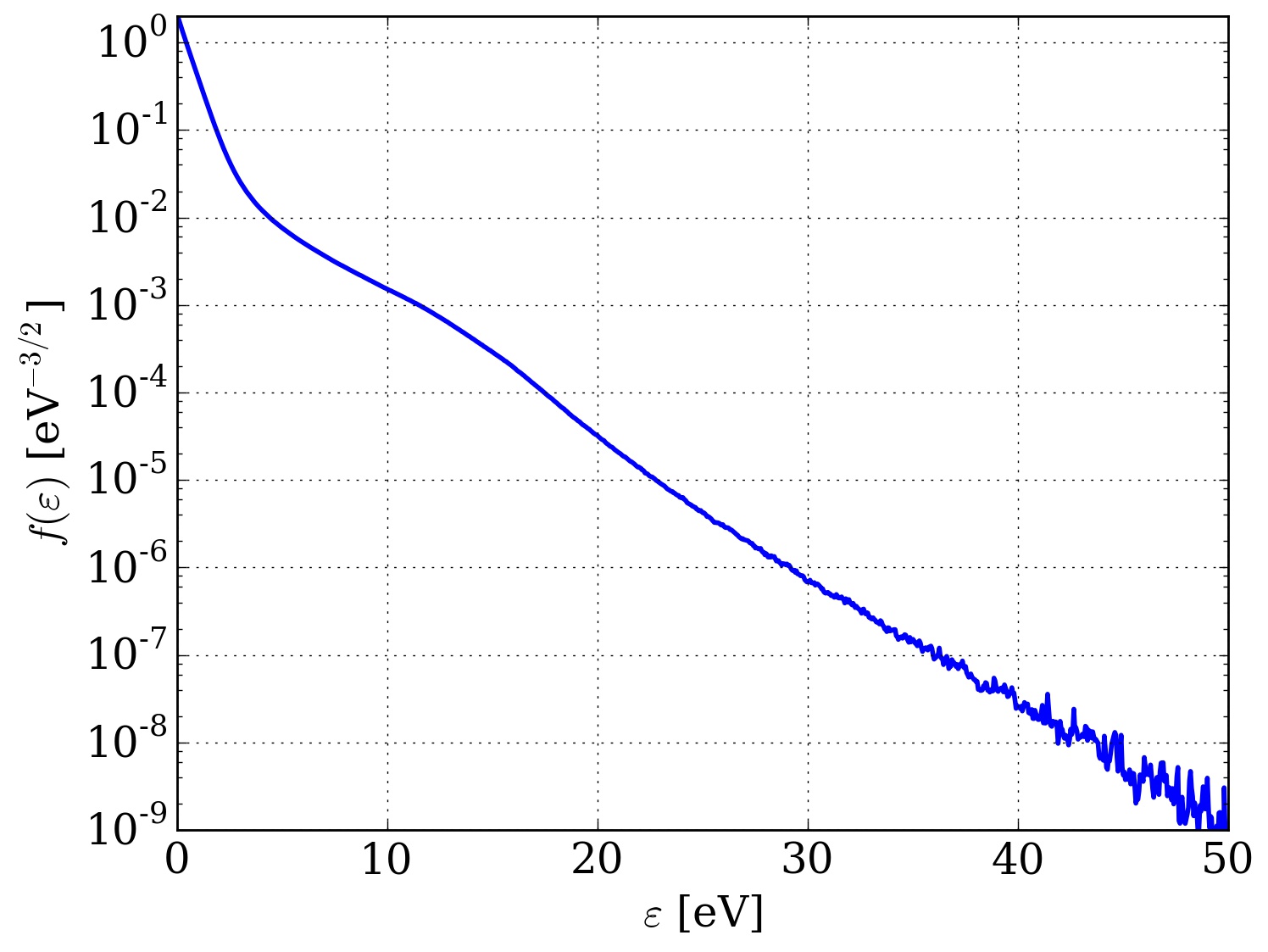} 
\caption{The temporally averaged electron energy probability function (EEPF) in the center of the discharge normalized as $\int f(\varepsilon) \sqrt{\varepsilon} {\rm d}\varepsilon = 1$ (data file: \texttt{eepf.dat}). Discharge conditions: argon $p=10$ Pa, $L=25$ mm, $T_{\rm{g}} = 350$~K, $V_{\rm{}} = 250$ V and $f_{\rm{}} = 13.56$ MHz. The number of energy bins is \texttt{N\_EEPF = 2000} and the resolution is \texttt{DE\_EEPF = 0.05} (eV).}
\label{fig:EDF10Pa}
\end{center}
\end{figure}

\begin{figure}[h!]
\centering
\begin{center}
\includegraphics[width=0.4\textwidth]{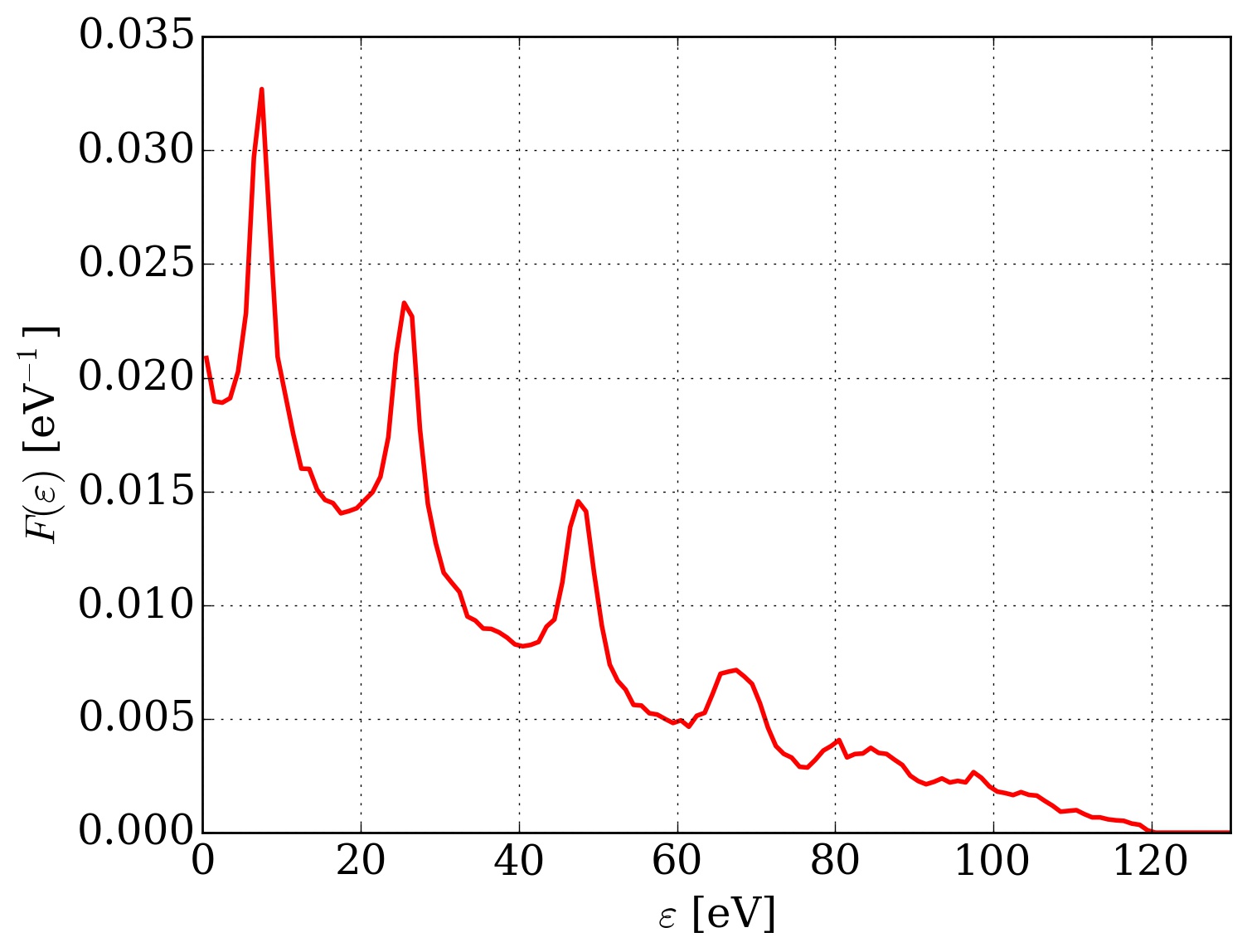} 
\caption{The temporally averaged flux-energy distribution (IFEDF) of the Ar$^+$ ions at the powered electrode normalized as $\int F(\varepsilon) {\rm d}\varepsilon = 1$ (data file: \texttt{ifed.dat}). Discharge conditions: argon $p=10$ Pa, $L=25$ mm, $T_{\rm{g}} = 350$ K, $V_{\rm{}} = 250$~V and $f_{\rm{}} = 13.56$ MHz. The number of energy bins is \texttt{N\_EEPF = 200} and the resolution is \texttt{DE\_EEPF = 1} (eV).}
\label{fig:IDF10Pa}
\end{center}
\end{figure}

More information about the energy distribution of the electrons (via the EEPF, $f(\varepsilon)$, ``measured'' at the central region of the plasma) and the ions (via their flux energy distribution, IFEDF, $F(\varepsilon)$, at the electrodes) is depicted in figures \ref{fig:EDF10Pa} and \ref{fig:IDF10Pa}, respectively. The EEPF reveals a behavior that is expected in Ar CCPs at low pressures \cite{godyak1990abnormally}. (In CCPs, the EEPF is known to exhibit a complex variation with position and time within the RF cycle. The average does not convey information about these changes, the investigation of which remains as future work of those readers who implement a straightforward extension of the eduPIC code to reveal this information.)

The IFEDF exhibits a series of peaks, characteristics for ion transport through the sheaths in the presence of charge-exchange collisions \cite{wild1991ion,schungel2017simple}. 

\subsection{Ion properties and ionization dynamics in dual-frequency discharges}

\label{sec:r3}

As it has already been mentioned in section \ref{sec:intro}, the control of the ion properties, viz. their flux and mean energy is of utmost importance for controlling surface reactions in plasma processing applications \cite{Tahsin,bogdanova2020virtual,kitajima2000functional}. Among the several approaches, the case of a voltage waveform that consists of two components with largely differing frequencies, i.e. the ``classical Dual-Frequency'' (DF) excitation in geometrically symmetric reactors can readily be studied with our simple PIC/MCC code as the DC (Direct Current) self-bias voltage in such systems is negligible \cite{Czarnetzki_2011} and its self-consistent computation can therefore safely be omitted. Below, we show representative simulation results for argon discharges excited by waveforms
\begin{equation}
    V(t) = V_{\rm HF} \cos(2 \pi V_{\rm HF} t) + V_{\rm LF} \cos(2 \pi V_{\rm LF} t),
\end{equation}
where the lower frequency (LF), $f_{\rm LF}$, is fixed at 2~MHz, and the higher frequency (HF), $f_{\rm HF}$, is in the range of 10--40~MHz. The respective voltage amplitudes, $V_{\rm LF}$ and $V_{\rm HF}$, and the additional discharge and simulation parameters are specified in table \ref{tab:table3}. We investigate two cases, with $p$ = 20 Pa (CASE 1) and 3 Pa (CASE 2). 

\begin{table} [ht]
\centering%
\caption{Set of parameters for the dual-frequency discharge simulations.}
\resizebox{\columnwidth}{!}{\begin{tabular}{|l|c|c|c|c|c|}
\hline
$V_{\rm LF}$ & 0 V & 100 V & 200 V & 300 V & 400 V \\ 
\hline
$L$ & \multicolumn{5}{ c| }{25 mm} \\
$T_{\rm g}$ & \multicolumn{5}{ c| }{350 K} \\
$f_{\rm LF}$ & \multicolumn{5}{ c| }{2 MHz} \\
\hline
{} & \multicolumn{5}{ c| }{\bf CASE 1} \\
$p$ & \multicolumn{5}{ c| }{20 Pa} \\
\hline
$f_{\rm HF}$ / $V_{\rm HF}$ & \multicolumn{5}{ c| }{10 MHz / 500 V} \\
\cline{2-6}
$W$             &  1.1$\times10^5$     & 1.0$\times10^5$      &   8.2$\times10^4$    &   6.1$\times10^4$    &   4.2$\times10^4$\\
$N_{\rm g}$     &  600                  & 600                   &   600                 &   600                 &   600\\
$N_{\rm t}$     &  31100                & 31200                 &   31400               &   31700               &   33200\\
\hline
$f_{\rm HF}$ / $V_{\rm HF}$ & \multicolumn{5}{ c| }{20 MHz / 250 V} \\
\cline{2-6}
$W$             &  1.7$\times10^5$     & 1.2$\times10^5$      &   9.2$\times10^4$    &   7.8$\times10^4$    &   6.8$\times10^4$\\
$N_{\rm g}$     &  600                  & 500                   &   500                 &   500                 &   500\\
$N_{\rm t}$     &  30700                & 25800                 &   26100               &   26200               &   26400\\
\hline 
{} & \multicolumn{5}{ c| }{\bf CASE 2} \\
$p$ & \multicolumn{5}{ c| }{3 Pa} \\
\hline
$f_{\rm HF}$ / $V_{\rm HF}$ & \multicolumn{5}{ c| }{30 MHz / 250 V} \\
\cline{2-6}
$W$             &  1.6$\times10^5$     & 1.5$\times10^5$      &   1.5$\times10^5$    &   5.5$\times10^4$    &   1.9$\times10^4$\\
$N_{\rm g}$     &  500                  & 500                   &   500                 &   500                 &   400\\
$N_{\rm t}$     &  30800                & 30900                 &   31300               &   31900               &   27200\\
\hline
$f_{\rm HF}$ / $V_{\rm HF}$ & \multicolumn{5}{ c| }{40 MHz / 150 V} \\
\cline{2-6}
$W$             &  1.5$\times10^5$     & 1.3$\times10^5$      &   1.1$\times10^5$    &   8.8$\times10^4$    &   2.8$\times10^4$\\
$N_{\rm g}$     &  500                  & 500                   &   500                 &   500                 &   400\\
$N_{\rm t}$     &  30700                & 30900                 &   31100               &   31300               &   26300\\
\hline
\end{tabular}}
\label{tab:table3}
\end{table}

\subsubsection{CASE 1: $p=20$ Pa}\hfill\\

\begin{figure*}[h!]
\centering
\begin{center}
\includegraphics[width=1.0\textwidth]{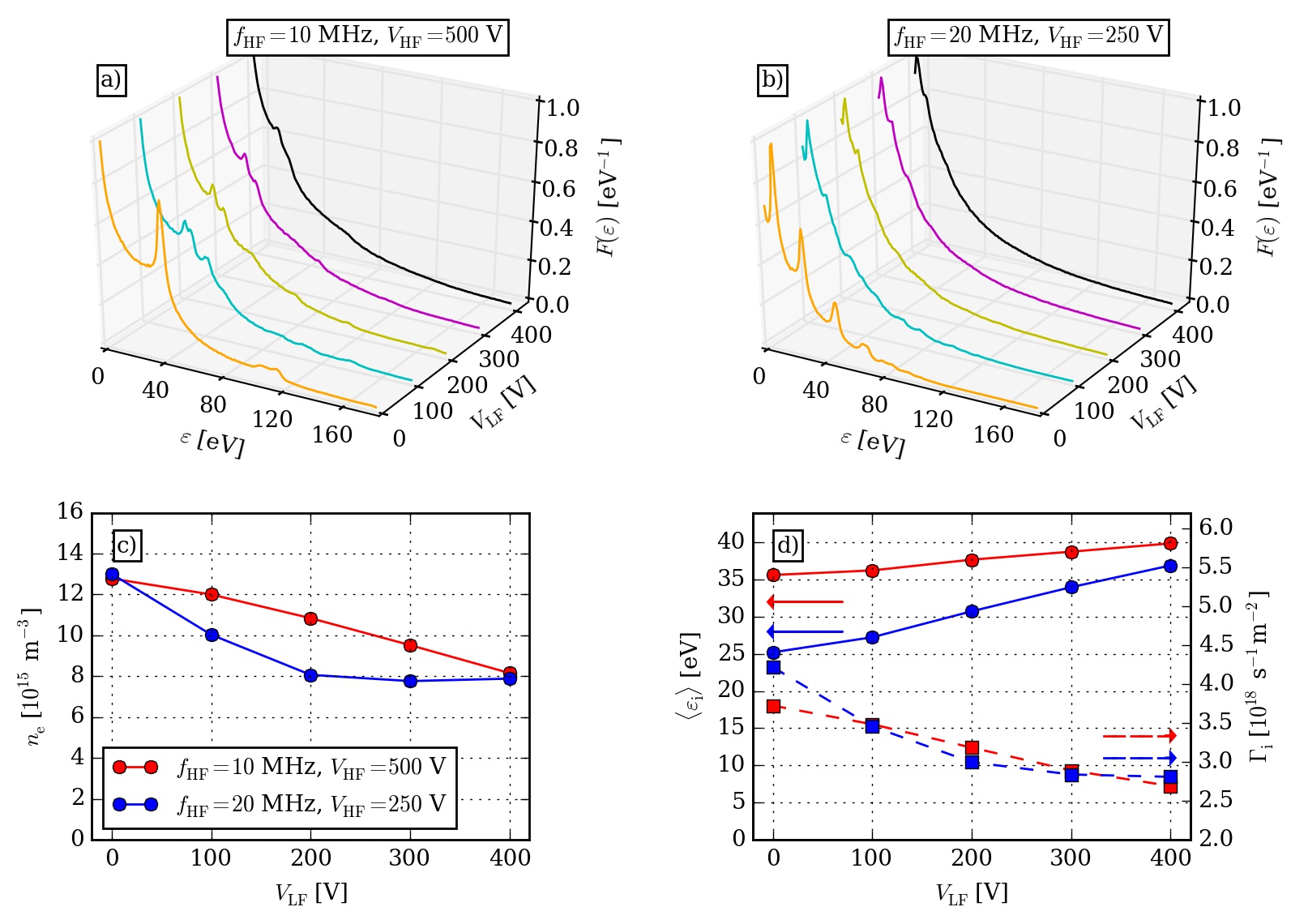} \caption{(a-b) IFEDFs as a function of the low frequency voltage for (a) $f_{\rm HF} = 10$ MHz, $V_{\rm HF} = 500$ V and (b) $f_{\rm HF} = 20$ MHz, $V_{\rm HF} = 250$ V. For a better comparison, the IFEDFs are normalized in a way that the maximum value is 1. (c) Peak electron density as a function of the low frequency voltage. (d) Mean ion energy at the powered electrode (solid lines, left axis) and the mean ion flux at the same electrode (dashed lines, right axis) as a function of the low frequency voltage. The colors of the lines in this panel correspond to those in panel (c). Discharge conditions: argon $p=20$ Pa, $L=25$ mm, $T_{\rm{g}} = 350$ K and $f_{\rm{LF}} = 2$ MHz. }
\label{fig:density20Pa}
\end{center}
\end{figure*}

\begin{figure*}[h!]
\centering
\begin{center}
\includegraphics[width=0.8\textwidth]{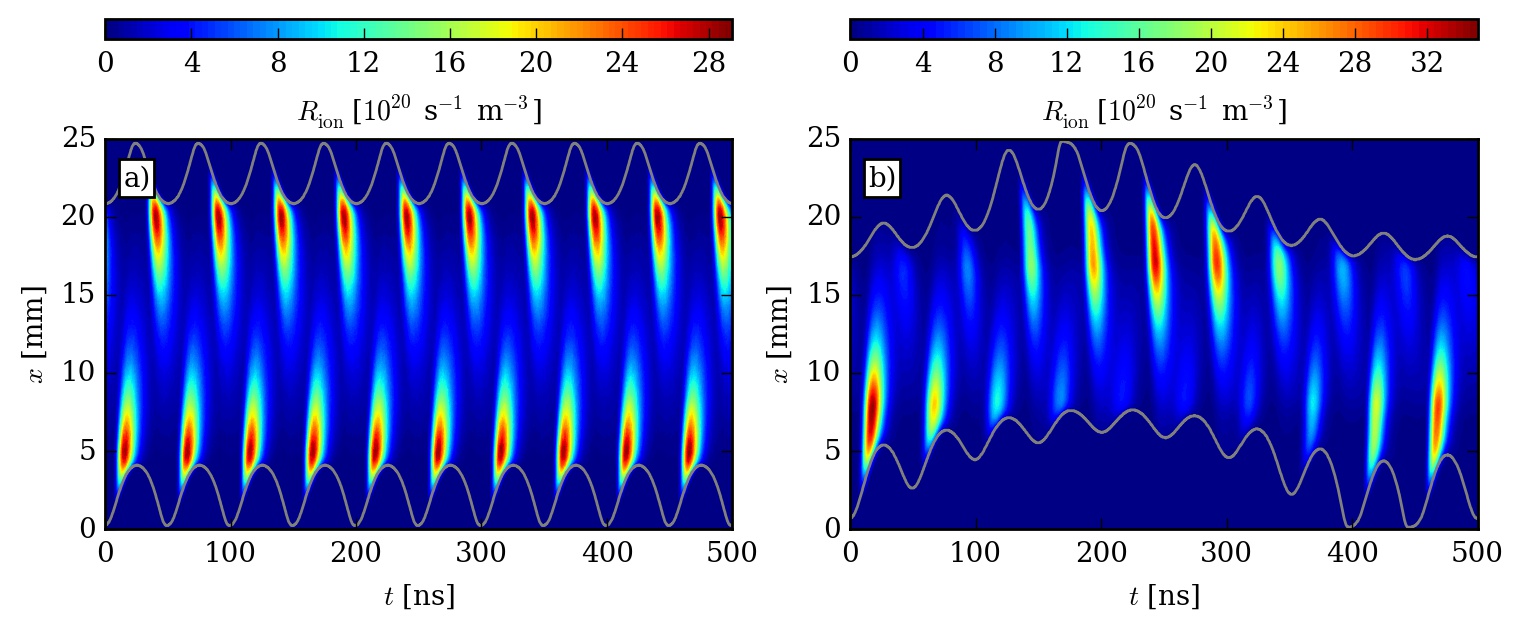} \caption{The spatial and temporal dynamics of the ionization rate for $f_{\rm{HF}} = 20$ MHz and $V_{\rm{HF}} = 250$ V. The low frequency voltage is $V_{\rm{LF}} = 0$ V in panel (a) and $V_{\rm{LF}} = 400$ V in panel (b). The gray solid lines represent the sheath edges. Discharge conditions: argon $p=20$ Pa, $L=25$ mm, $T_{\rm{g}} = 350$ K and $f_{\rm{LF}} = 2$ MHz. The time shown on the horizontal axis corresponds to one period of the LF excitation, $T_{\rm LF}=1/f_{\rm LF}$.}
\label{fig:ionization20Pa}
\end{center}
\end{figure*}

\noindent We start the presentation of the results with the $p$ = 20 Pa case, for which we usd $f_{\rm HF}$ values of 10 MHz and 20 MHz, and $V_{\rm HF}$  values of 500~V and 250~V, respectively. Figure \ref{fig:density20Pa}(a) and (b) present the flux-energy distributions of the Ar$^+$ ions ($F(\varepsilon)$), for these values of $V_{\rm HF}$, along with a series of $V_{\rm LF}$ values. At $V_{\rm LF} =0$ (corresponding to a single-frequency excitation) the $F(\varepsilon)$ function peaks at low energy. This is a result of the collisional transport of the Ar$^+$ ions through the sheaths due to their mean free path being significantly shorter than the sheath length at this relatively high pressure. The ionization source function corresponding to the single-frequency case with $f_{\rm{HF}} = 20$ MHz and $V_{\rm{HF}} = 250$ V is shown in figure \ref{fig:ionization20Pa}(a). Ionization maxima in the vicinity of the expanding sheath edges can be seen to be repeated as dictated by the value of the high frequency. 
The application of a low frequency voltage results in a decay of the plasma density as it can be seen in figure \ref{fig:density20Pa}(c). As a consequence of this, the ion flux to the electrodes also decreases with increasing $V_{\rm LF}$, as shown in figure \ref{fig:density20Pa}(d). The same figure also reveals that the mean energy of the ions hardly increases as a function of $V_{\rm LF}$, which, at least in principle, should serve as the ``control parameter'' for the mean ion energy and is supposed to have little influence on the ion flux. The results shown here confirm that this is not the case, i.e. the idea of using DF excitation for an independent control of the ion properties does not work efficiently under the conditions of CASE 1.

The decay of the plasma density and the ion flux can be attributed to the changes of the spatio-temporal distribution of the ionization rate, $R_{\rm ion}$ as a function of $V_{\rm LF}$. An example for $R_{\rm ion}$, obtained at $V_{\rm LF}$ = 400~V (at $f_{\rm{HF}} = 20$ MHz and $V_{\rm{HF}} = 250$ V) is displayed in figure \ref{fig:ionization20Pa}(b). We observe a complicated variation of the sheath length, being modulated by both the low and high frequency voltages. As a result, (i) a more intensive ionization is observed as compared to the single-frequency case (figure \ref{fig:ionization20Pa}(a)) when the sheath expands faster near the electrodes when the two applied voltage waveforms  ``interfere'' constructively, and (ii) a significantly lower ionization rate is obtained when the sheath length is large and the high frequency oscillations of the sheath length occur in the domain of higher ion density. The second effect is more pronounced and results in a decrease of the ionization rate, and consequently, in a decreased plasma density at $V_{\rm LF} > 0$. This interplay between the two driving voltage components is called as the frequency coupling effect \cite{schulze2007space,waskoenig2010nonlinear}. Due to their increased length, the sheaths become even more collisional and the increased total voltage can just about compensate for the higher collisional energy losses of the ions, and results in about the same mean ion energy at any $V_{\rm LF} > 0$. 
These observations hold for $f_{\rm{HF}} = 20$ MHz as well, the mean energy of the ions increases only slightly with $V_{\rm LF}$ and the flux-energy distributions shown in figure \ref{fig:density20Pa}(b) are rather similar at any $V_{\rm LF}$, except for the disappearance of the peak created by charge transfer collisions \cite{wild1991ion}. We recall that the origin of these peaks is that ions that have undergone charge transfer collisions at times of a low sheath voltage can accumulate in certain spatial regions and get accelerated again when the sheath voltage reaches an appreciable value \cite{schungel2017simple}. The periodicity of the decay and rise of the sheath electric field under single-frequency excitation can this way synchronize the motion of many ions, forming peaks in $F(\varepsilon)$. At DF excitation, however, the sheath voltage is low only during very limited times during the LF radio-frequency cycle, making such a synchronized ion motion hardly possible, which leads to the disappearance of the peaks. 

\subsubsection{CASE 2: $p=$ 3 Pa}\hfill\\

\noindent At this pressure, high-frequency values of $f_{\rm HF}$ = 30 MHz and 40 MHz values are used. These higher values of the frequency, as compared to those used in Case 1, are needed to establish sufficient ionization at the significantly lower value of the pressure considered now. In order to keep the plasma density near the same value as in Case 1, $V_{\rm HF}$ values of 250~V and 150~V are used, respectively, for the $f_{\rm HF}$ values given in table \ref{tab:table3}. 

\begin{figure*}[h!]
\centering
\begin{center}
\includegraphics[width=1.0\textwidth]{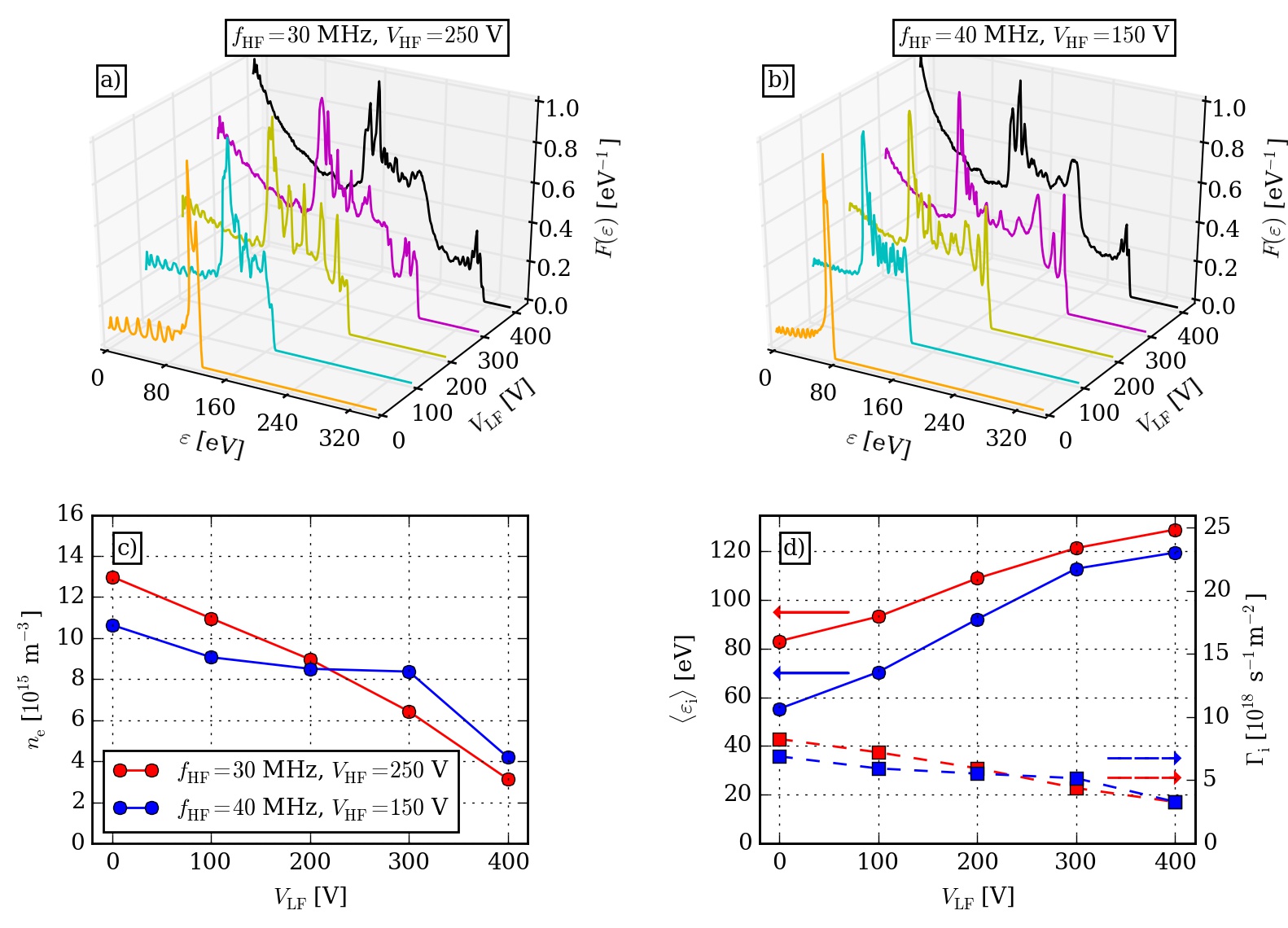}
\caption{(a-b) IFEDFs as a function of the low frequency voltage for (a) $f_{\rm HF} = 30$ MHz, $V_{\rm HF} = 250$ V and (b) $f_{\rm HF} = 40$ MHz, $V_{\rm HF} = 150$ V. For a better comparison, the IFEDFs are normalized in a way that the maximum value is 1. (c) Peak electron density as a function of the low frequency voltage. (d) The colors of the lines in this panel correspond to those in panel (c). The mean ion energy at the powered electrode (solid lines, left axis) and the mean ion flux at the same electrode (dashed lines, right axis) as a function of the low frequency voltage. Discharge conditions: argon $p=3$ Pa, $L=25$ mm, $T_{\rm{g}} = 350$ K and $f_{\rm{LF}} = 2$ MHz.}
\label{fig:density3Pa}
\end{center}
\end{figure*}

\begin{figure*}[h!]
\centering
\begin{center}
\includegraphics[width=0.8\textwidth]{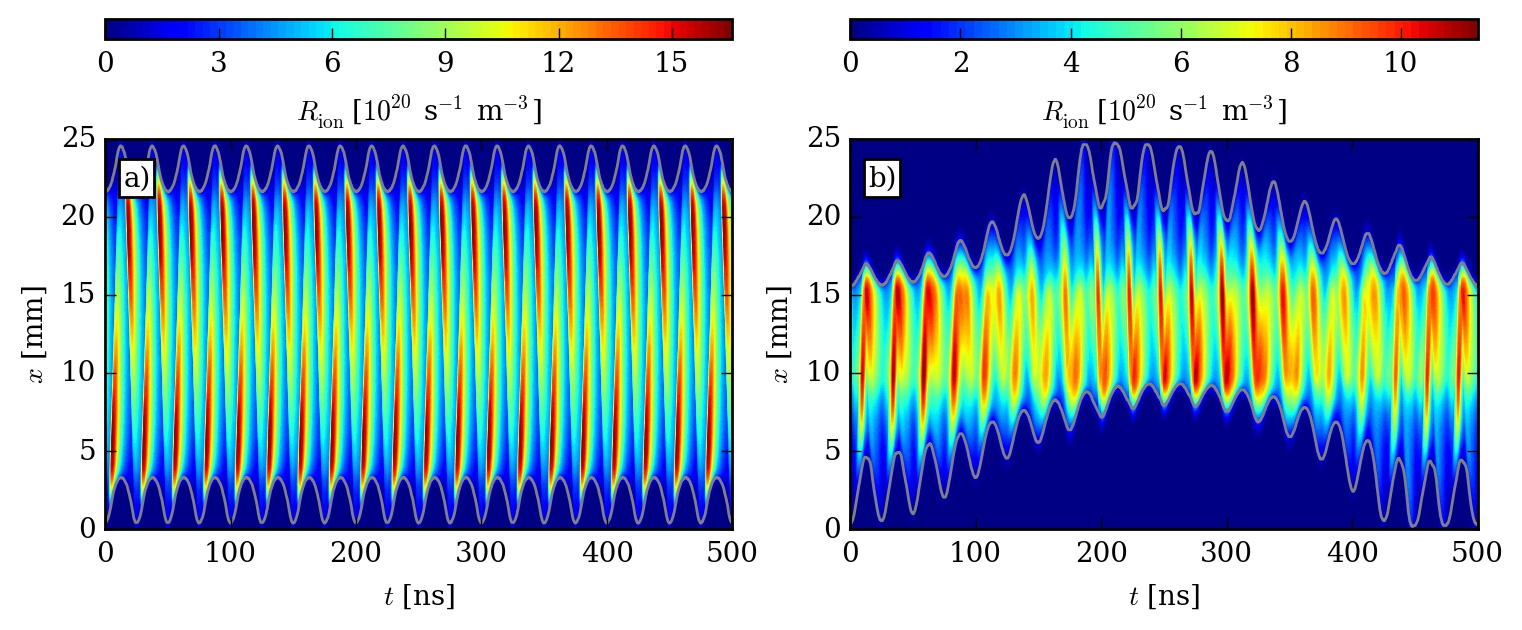}
\caption{The spatial and temporal dynamics of the ionization rate for $f_{\rm HF} = 40$ MHz and $V_{\rm HF} = 150$ V. The low frequency voltage is $V_{\rm LF} = 0$ V in panel (a) and $V_{\rm LF} = 400$ V in panel (b). The gray solid lines represent the sheath edges. Discharge conditions: argon $p=3$ Pa, $L=25$ mm, $T_{\rm{g}} = 350$ K and $f_{\rm{LF}} = 2$ MHz. The time shown on the horizontal axis corresponds to one period of the LF excitation, $T_{\rm LF}=1/f_{\rm LF}$.}
\label{fig:ionization3Pa}
\end{center}
\end{figure*}

The flux-energy distributions of the Ar$^+$ ions are shown in figure \ref{fig:density3Pa}(a) and (b), respectively, for $f_{\rm HF}$ = 30 MHz and 40 MHz. In the single-frequency cases ($V_{\rm LF}=0$ V), the observed IFEDFs exhibit a prominent peak at high energy (that corresponds to the time-averaged sheath voltage) indicating a nearly collisionless transfer of the positive ions through the sheaths. With the application of the low-frequency excitation the IFEDFs change in a complicated manner. The most important change is that these functions extend towards higher energy with increasing $V_{\rm LF}$, is rather obvious. The increased mean ion energy is also confirmed by the data shown in figure \ref{fig:density3Pa}(d). The simultaneous increase of the low-energy part in the $F(\varepsilon)$ functions in figure \ref{fig:density3Pa}(a,b), is however, an indication of the increase of the sheath length that was also observed at $p$ = 20 Pa, discussed earlier. The increase of the sheath length is clearly demonstrated by the plot of the ionization rate, $R_{\rm ion}$, in figure \ref{fig:ionization3Pa}(b). Due to the lower pressure, the maximum of the ionization rate moves towards the center of the plasma as $V_{\rm LF}$ is increased, at otherwise same conditions. The coupling of the excitation harmonics results in this case as well in a decrease of the plasma density (figure \ref{fig:density3Pa}(c)) and the ion flux (figure \ref{fig:density3Pa}(d) with increasing $V_{\rm LF}$ in the absence of secondary electron emission from the electrodes \cite{donko2010effect}. Nonetheless, the mean ion energy increases significantly and can be tuned by the control parameter $V_{\rm LF}$ over an appreciable range, unlike in the case of higher pressures. 

\section{Summary and ideas for the further development of the code}
\label{sec:summ}

In this work, we have described a simple Particle-In-Cell/Monte Carlo Collisions code for 1D electrostatic simulations of capacitively coupled radiofrequency plasmas. Following an introduction to the basics of the approach, we provided comprehensive information about the structure of the eduPIC code, its compilation and execution, as well as about the content of the data files created. The operation of the code was explained at the level of algorithmic details. Subsequently, a set of reference results was presented, with the default parameter settings of the code. When downloaded and properly executed by a reader, the code should reproduce these results without any modifications of the code and the simulation settings. We also illustrated the capabilities of this basic code by presenting some simulation results for dual-frequency CCPs. 

As it has already been emphasized, this code intends to be a ``starting tool'' that can be optimized in many respects (e.g. in handling the collisions~\cite{Nanbu2000,sphere}) and can be developed further to address various research questions that the reader of this work may have. To assist such activities, we provide our code as an open source item, via Github, in three computing languages: the basic (C) version is complemented with more sophisticated versions available in C\texttt{++} and Rust \cite{eduPIC_SOURCEs}. 

Below we give several suggestions for the further development of the code, which will greatly widen its scope of applicability and/or enhance its performance. 

\begin{itemize}
    \item {\it Implementation of the emission of ion-induced secondary electrons from the electrodes.} In a stochastic approach the ion-induced secondary electron yield, $\gamma$, is understood as a probability: whenever an ion reaches the surface of an electrode, an electron is emitted with this probability (as long as $\gamma<1$), i.e., when $R_{01} < \gamma$. The simplest approach is to use a constant $\gamma$, in a more sophisticated model, an energy-dependent $\gamma$ can be adopted. For data for Ar$^+$ see \cite{phelps1999cold}, as well as \cite{phelps81999plasma} for  corrections to some of the formulas given in \cite{phelps1999cold}.
    
    \item {\it Implementation of the elastic reflection of electrons at the electrodes.} Whenever an electron reaches the surface of an electrode, the electron is reflected elastically with a probability $\eta_{e}$. In the simplest approach, the probability of the elastic reflection can be set to a constant value (e.g. $\eta_{e}=0.2$ \cite{Kollath}, independently of the discharge conditions and surface properties). In a more complex approach, $\eta_{e}$ is the function of the energy and angle of incidence of the electron and depends on the surface properties \cite{braginsky2011experimental,horvath2017eSEEmodel}. (One can go even beyond considering elastic reflection and include as well inelastic reflection and secondary electron emission due to electron impact \cite{horvath2017eSEEmodel}.)
    
    \item {\it Implementation of the computation of the self-bias voltage that develops when the discharge is driven by tailored waveforms.} Tailored excitation voltage waveforms provide a possibility to control the energy distribution of the ions (and electrons) reaching the electrodes (see e.g. \cite{Lafleur_2015}). The DC self-bias voltage can be computed by monitoring the currents of the oppositely charged particles at the electrodes. At each electrode, these currents have to compensate each other on time average, and the self-bias voltage can be adjusted in the simulation in an iteration cycle to achieve this balance. (The time-averaging should cover a number of RF cycles so that the statistical fluctuations of the particle currents should be sufficiently suppressed.) 
    
    \item { {\it Implementation of different electrode materials at the powered and grounded electrode.} Electrodes made of different materials can be modeled by using different probabilities for the ion-induced secondary electron emission ($\gamma$) and/or the elastic reflection of electrons ($\eta_e$) at both electrodes. The DC self-bias voltage due to the secondary electron induced asymmetry of the discharge \cite{lafleur2013secondary} and/or the asymmetry induced by electron reflection \cite{Korolov_2016_esticking} can be determined as suggested above.}
    
    \item {\it Improvement of the scattering model by incorporating anisotropic scattering for electron-atom collisions.} At high electron energy the scattering may be forward-peaked, which can be incorporated into the code via the proper differential cross sections. For guidance, see e.g., \cite{okhrimovskyy2002electron,janssen2016evaluation}. 
    
    \item {\it Calculation of the position of the sheath edge.} For the definition of the sheath edge see \cite{brinkmann2007beyond}. Whenever a quasineutral region of the plasma exists, the procedure described in this reference gives a guidance for deriving the (time-dependent) position of the sheath edge from the (time-dependent) electron and ion density profiles. When the sheath length is known, further important information can be obtained from the existing data: (i) the voltage drop over the sheath can be found from the spatio-temporal distribution of the potential, and (ii) the net charge inside the sheath can be found from the density distributions of the charged particles. 
    
    \item {\it Change the code into a cylindrical version.} By considering a long cylindrical setup with coaxial electrodes\cite{Birdsall_2004,verboncoeur1993simultaneous,wilczek2018disparity}, plasma sources with unequal electrode surface areas can be simulated with a 1D approach. In this case, plasma parameters change as a function of the radial coordinate only. The computation of the charged particle densities, the solution of the Poisson equation and the integration of the equation of motion need modifications. In such a system, a self-bias voltage develops due to the unequal electrode surfaces. This voltage can be computed as described in one of the points above.
    
    \item {\it Include an external circuit.} Real plasma systems are driven by generators and matching networks to optimize power coupling into the plasma. In \cite{verboncoeur1993simultaneous}, a detailed description of the simultaneous solution of the Poisson equation and the equation for an external electrical circuit is provided, for various 1D geometries. 

    \item {\it Introduce the null-collision approach to select colliding particles.} Instead of evaluating the collision probability using eq. (\ref{eq:pcoll}) that involves significant computation, for each particle, a certain number of particles can be assigned to undergo collisions, based on the maximum of  the collision frequency of the given species. Using this approach the set of processes is expanded with the ``null collision'' \cite{Skullerud_1968,vahedi1995monte}  - upon the occurrence of this, the particle proceeds without any change. The method significantly increases the efficiency of selecting colliding particles.
    
    \item {\it Implement an iteration algorithm that allows using the discharge power as an input quantity.} The computation of the discharge power is implemented in the voltage-driven eduPIC code. Try to implement an iteration cycle that adjusts the voltage automatically in order to reach the power level specified. This ``control loop'' has to have sufficiently low gain to prevent diverging oscillations of the voltage as a result of its adjustment based on the difference between the specified and actual power levels. The latter should be measured over a sufficiently high number of RF cycles to provide an accurate measurement of the actual power. 
    
    \item {\it Parallelization of the code.} Modern computer architectures provide access to various paradigms of parallel code execution. In order to fully utilize the compute capacity of the hardware, code parallelization is a key technique. At the lowest level, the vectorization of simple operations repeated on consecutive data elements is performed by modern compilers by default. For multi-thread (multi CPU core) execution several methods exist, like OpenMP~\cite{dagum1998openmp}, pthreads~\cite{pthread}, or those provided by recent generations of the C\texttt{++} standard library. However, the PIC/MCC method is difficult to parallelize efficiently, as the Monte Carlo type of collision handling incorporates a high degree of code branching and unpredictable execution orders due to the heavy use of random numbers. A careful redesign of data structures and execution loops is necessary to optimize parallel performance. Alternatively, the application of massively parallel Graphics Processing Units (GPUs) represent attractive architectures for code acceleration, as discussed in~\cite{Juhasz2020}.

    \end{itemize}

Apart from the above list of ideas and suggestions, there is a number of emerging topics that deserve interest and can be studied in CCPs via extending the eduPIC code, e.g. the investigation of the effects of magnetic fields on the plasma and various asymmetry effects (e.g. \cite{wang2020electron,skarphedinsson2020tailored}). 
There is also a possibility to modify the eduPIC code to be able to describe various plasma sources, e.g., pulsed discharges. The exploration and the discussion of these possibilities goes, however, beyond the scope of this work.

As a final remark, it should be kept in mind that the code and its possible extensions are always built on {\it discharge models}, which always represent a simplification of real systems. Therefore, one cannot expect simulations to perfectly reproduce the physics of the real systems. The expected accuracy greatly depends on the completeness of the model. Also, following any modifications of the code, it is a good practice to benchmark the results with those of independent studies reported in the literature, if available. 

\section*{Disclaimer}

The eduPIC (educational Particle-in-Cell/Monte Carlo Collisions simulation code), Copyright $\copyright$ 2021  Zolt\'an Donk\'o \emph{et al.} is free software: you can redistribute it and/or modify it under the terms of the GNU General Public License as published by the Free Software Foundation, version 3.    This program is distributed in the hope that it will be useful, but WITHOUT ANY WARRANTY; without even the implied warranty of
MERCHANTABILITY or FITNESS FOR A PARTICULAR PURPOSE.  See the GNU General Public License for more details at \url{https://www.gnu.org/licenses/gpl-3.0.html}. 
    
\ack This work was supported by the National Office for Research, Development and Innovation (NKFIH) of Hungary via the grants K-134462 and FK-128924, by the German Research Foundation in the frame of the project, ``Electron heating in capacitive RF plasmas based on moments of the Boltzmann equation: from fundamental understanding to knowledge based process control'' (No. 428942393). MV was supported by the \'UNKP 20-3 New National Excellence Program of the Hungarian Ministry for Innovation and Technology from the source of the National Research, Development and Innovation Fund. ZD gratefully acknowledges discussions with L. C. Pitchford on electron scattering models and thanks M. J. Kushner and F. J. Schulze for their useful comments and advice related to the manuscript.

\section*{References}
\bibliography{eduPIC}
\bibliographystyle{iopart-num}

\end{document}